\documentclass[twocolumn]{aastex7}
\usepackage{graphicx}	
\usepackage{amsmath}	
\usepackage{orcidlink}

\usepackage{booktabs}

\shorttitle{miscibility and sub-Neptune demographics}

\usepackage{hyperref}
\usepackage{xcolor}

\begin{document}

\title{\fontsize{14}{18}\selectfont The Influences of Hydrogen-Silicate-Iron Miscibility on the \\[0.6ex] Demographics of Sub-Neptunes and Super-Earths}

\correspondingauthor{Edward D. Young}

\author[0000-0002-1299-0801]{Edward D. Young}
\affiliation{Department of Earth, Planetary, and Space Sciences, University of California, Los Angeles, CA 90095, USA}
\email[show]{eyoung@epss.ucla.edu}

\author[0009-0005-1133-7586]{Aaron Werlen}
\affiliation{Department of Earth, Planetary, and Space Sciences, University of California, Los Angeles, CA 90095, USA}
\email{werlen@ucla.edu}

\begin{abstract}

Models based on variable miscibility among hydrogen, molten silicate, and molten iron, coupled with atmospheric escape, can reproduce the observed 
occurrence density structure of 
sub-Neptunes and super-Earths in mass-radius space. The models are also consistent 
with the radius gap and the observed radius-period relationship exhibited by these planets. The degree of overlap between predicted and observed planetary occurrences suggests that 
hydrogen-silicate-iron miscibility may serve as a unifying concept for the 
formation and evolution of these planet classes. The well-defined equilibrium 
conditions at the boundary between supercritical magma oceans and the overlying hydrogen-rich 
envelopes are important features of the models.  
Planets formed with less than  $\sim1\%$  hydrogen by mass develop discrete, terrestrial-like metallic cores, while those accreting greater hydrogen concentrations are predicted to have 
fully miscible interiors and no discrete metal cores. Hydrogen-silicate-iron miscibility 
provides an overarching explanation for the full range of sub-Neptune and 
super-Earth architectures based on the accreted hydrogen 
mass fraction and the phase equilibria governing silicate, iron metal, and H$_2$ miscibility.    
\end{abstract}


\section{Introduction}
Sub-Neptunes are apparently the most abundant planet type in the Galaxy  \citep[e.g.,][]{fressin2013a, fulton2017a}. These planets are characterized by bulk densities intermediate between those of rocky planets and Neptune-like planets.  
Their low densities relative to compressed rock and metal suggest that sub-Neptunes may be mixtures of roughly equal mass fractions of rock and water \citep[e.g.,][]{Fortney2007, zeng2019a} or molten/partially molten interiors beneath hydrogen-rich primary atmospheres \citep{Chachan2018}. The H$_2$-rich envelopes in this interpretation constitute percent levels of the planet masses in order to match their bulk densities \citep[e.g.][]{Bean2021}.   

There is a rich history of attempting to infer the chemistry and structure of sub-Neptunes from their bulk densities. Applications of equations of state (EoSs) for silicate, iron metal, water, and H$_2$ have been used to invert mass and radius data for the compositions and mass fractions of various layers comprising individual planets.  These previous studies account for compression of rock and metal \citep[e.g.,][]{Seager2007, Valencia2006}, and inclusion of H$_2$ and water layers \citep[e.g.,][]{Sotin2007, RogersSeager2010, LopezFortney2014, Zeng2016, NixonMadhusudhan2021, Aguichine2021, Vazan2022}. Probabilistic approaches have been applied to make clear where degeneracies can and cannot be broken \citep[e.g.,][]{Wolfgang_2015, Dorn2017}. 

Dense H$_2$-rich atmospheres can limit cooling of sub-Neptunes such that initialy molten interiors stay molten for Gyr timescales \citep{Ikoma2006, LopezFortney2014, ginzburg2016a}.  As a result, there has been a recent focus on interaction between hydrogen atmospheres and underlying magma oceans \citep{Kite2019, Bean2021, Vazan2022, Schlichting_Young_2022, Misener2023, Charnoz2023, Young_2024, RogersYoungSchlichting2025_MNRAS, Young2025_Differentiation, werlen_atmospheric_2025, werlen_sub-neptunes_2025}. Experimental verification for the implications of H$_2$-silicate-metal interactions deduced from extrapolation of chemical thermodynamics and {\it ab-intio} molecular dynamics is just now becoming available \citep{Miozzi2025, Horn2025_WetPlanets}.

Super-Earths are evidently composed primarily of rock and metal.  They are most often thought to represent the stripped cores left behind by nearly complete escape of the primary atmospheres comprising sub-Neptunes \citep{owen2013a, owen2017a, LopezFortney2014, vaneylen2018a, gupta2019a}. Key features of these arguments are the existence of a radius valley, or gap, between super-Earths and sub-Neptunes that is near $1.8 M_{\oplus}$ further from the host star (longer orbital periods) and $2.0 M_{\oplus}$ nearer to the star (shorter orbital periods)  \citep{fulton2017a, vaneylen2018a, gupta2019a}.  An alternative model suggests that with an appropriately broad initial mass function for rocky/metal cores and accretion late in the evolution of a protoplanetary disk, the period-dependent radius gap can be a result of planet accretion \citep{Lee_2021}. 

A result of a deeper understanding of the physical chemistry of hydrogen, silicate, and metal interactions is that the traditional view of sub-Neptunes and their potential super-Earth descendants as having purely metal cores and silicate mantles, with or without surviving primary hydrogen-rich envelopes, may not be universally correct.  Rather, at relatively moderate temperatures and pressures in the context of these planets, H$_2$ and silicate become entirely miscible \citep{gilmore_core-envelope_2025}, with complete miscibility defining the boundary between a supercritical magma ocean and overlying H$_2$-rich envelope \citep{Young_2024}.  In addition, within the interior (sometimes referred to as the ``core" to distinguish it from the overlying envelope), H$_2$, silicate, and iron metal are predicted to become fully miscible \citep{Young2025_Differentiation}. 

In this paper we seek to answer the question of whether or not the newly recognized physical chemistry of hydrogen-silicate-metal planets is consistent with the populations of sub-Neptunes and super-Earths. The paper is organized as follows: our epistemological approach for constructing model planets for comparisons with observed planets is presented in \S~\ref{section:epsitemological_approach};
the planet models are described in \S~\ref{section:planet_models}; the random sampling
of planet models is explained in \S~\ref{section:monte_carlo_sampling} and the resulting model
planet population is compared with observed planets in \S~\ref{section:results};
the implications of the results are discussed in \S~\ref{discussion}; and our
conclusions are summarized in \S~\ref{section:conclusions}.

\section{Epistemological Approach}
\label{section:epsitemological_approach}
Two different approaches are often used when attempting to explain the observable properties of sub-Neptunes and super-Earths:   
(1) \emph{Forward evolutionary models}, in which a planet's thermal 
state and atmospheric mass fraction are evolved from specified initial conditions of specific entropy, mass, and mass fraction of volatile components  
\citep[e.g.,][]{LopezFortney2014, owen2017a}; and  
(2) \emph{Inverse statistical inference}, in which emphasis is placed on constraining present-day planetary 
properties necessary to explain the observables   
\citep[e.g.,][]{RogersSeager2010, Dorn2017, Wolfgang_2015}.  
These approaches differ not only in methodology and philosophy, but also in computational cost and sensitivity to inferred initial conditions.

A limitation of forward models is their computational cost. 
High-fidelity evolution models include detailed solutions for energy transfer, e.g., $dE_p/dt = -L_{\rm int}$, where $E_p$ is the total thermal and gravitational potential energy of the planet and $L_{\rm int}$ is the intrinsic net luminosity leaving the body, as well as hydrodynamic or energy-limited atmospheric escape. The forward integrations can require  
time-stepping through tens of thousands of model calls per planet, in some cases accelerated by using pre-computed grids or libraries that can be used for interpolation. 

Forward models place an emphasis on the time evolution from an initial state. However, these initial states are plagued by uncertainties surrounding the retention of heat during the accretion process \citep{Young_2024}. In addition, sub-Neptunes undergo rapid thermal evolution, losing most of their primordial heat within their first $10$ to $10^2$ Myr \citep{LopezFortney2014}. Although these planets cool rapidly at early times, evidence of their initial conditions can persist until
late in their evolution \citep[e.g.,][]{arevalo_sub-neptune_2025}. However,
these radii differences appear to dwindle to a few percent at very
late ages, and disappear entirely by ~10 Gyr. Conversely, while forward models attempt to determine  what planet results from a set of inferred initial conditions, inverse methods instead place the emphasis on the extant, underlying chemistry and structure consistent with observable features like mass and radius.  

Primarily because of the computational costs associated with forward evolution models in the present case, but also because of uncertainties surrounding the completeness of initial conditions, we adopt an inverse approach in this paper. Because evolutionary pathways are not uniquely recoverable from present-day 
observables, the inverse problem is fundamentally probabilistic. The goal is to infer the minimum number of plausible initial conditions and evolutionary histories that are compatible with the present-day observables. This approach minimizes assumptions about initial conditions, e.g., total specific entropy that sets the central core temperature but is effectively erased on 100 Myr timescales, and computational costs, which are limiting in this application. 

Our approach is to draw random values for variables that prescribe the states of sub-Neptune planets that include, where appropriate, complete miscibility in their interiors, using a range for each variable that is consistent with the likely ages matching observed planet ages.  We use those variables to construct model populations of sub-Neptunes. We then compute the probability that each model planet could have retained its primordial H$_2$-rich atmosphere given the host-star environment and age assigned to it.  If the probability is low, then this planet is stripped of its primary atmosphere, leaving behind a core that is now a super-Earth. We compare the resulting demographics of model planets to the observed population of super-Earths and sub-Neptunes. 

\section{Planet Models}
\label{section:planet_models}

\subsection{Planet Structure}
Many of the methods for calculating the planet models used here have been described previously \citep{Young2025_Differentiation, RogersYoungSchlichting2025_MNRAS}. However, there are some modifications to the prior published work, and so the models are explained here. To derive the core\footnote{Throughout this paper we will follow the convention in the astronomical fields and use the term ``core" to mean everything interior to the envelope or atmosphere of the planet.  In some cases, modifieres like ``metal" will be used to signify a separate phase in the interior, consistent with geophysical usage. }  we solve the system of equations \citep[e.g.,][]{Seager2007}:
\begin{equation}
    \frac{dm}{dr}=4 \pi r^2 \rho,
\label{eq:dmdr}
\end{equation}
\begin{equation}
    \frac{dP}{dr}=-\frac{Gm \rho}{r^2},
\label{eq:dPdr}
\end{equation}
and,
\begin{equation}
\left(\frac{dT}{dr}\right)_S
= -\,\frac{\gamma(\eta)\,T}{K_S(\eta)}\,\rho\,g
\label{eq:dTdr}
\end{equation}
where $m$ is the mass contained within radius $r$, $\rho$ is mass density, $\gamma(\eta)$ is the Gr\"uneisen parameter that varies with $P$ and $T$ through the compressibility factor  $\eta = \rho/\rho_0$, and $K_S(\eta)$ is the isentropic bulk modulus.  Here one assumes the core is fully convective, with a limiting isentropic $dT/dP$ gradient. In practice, we evaluate the bulk modulus using  $K_S(\eta)\equiv \eta\,d\!\left(P\!\left(\eta,T_S\right)\right)/d\eta$ where $P\!\left(\eta,T_S\right)$ is the total isentropic pressure along the adiabat with temperature $T_S$.    
Numerically integrating Equations (\ref{eq:dmdr}) through (\ref{eq:dTdr}) through one or more layers (with or without a metal core) yields a density and temperature profile for the planet interior. 

Our use of the adiabat is predicated on a molten interior throughout. We estimate the depression of the MgSiO$_3$ solidus by dissolved H$_2$
using the ideal cryoscopic relation
\begin{equation}
\Delta T_m =
-\frac{R T_m^2}{\Delta H_{\rm fus}}\ln x_{\rm MgSiO_3},
\label{eqn:cryo}
\end{equation}
where $x_{\rm MgSiO_3}$ is the mole fraction of MgSiO$_3$ in the melt.
For representative binodal conditions near 4~GPa, an H$_2$ content of
$\sim 2$~wt\% corresponds to $x_{\rm MgSiO_3}\simeq 0.50$. Taking
$T_m\simeq 3200$~K and $\Delta H_{\rm fus}\simeq 75$~kJ~mol$^{-1}$
\citep{Stixrude2005, Stixrude2014} gives
$\Delta T_m\simeq 800$~K. Because
$\Delta T_m \propto T_m^2/\Delta H_{\rm fus}$ and $T_m$ increases
strongly with pressure, the melting-point depression is expected to
remain large, and likely increase, deeper in the planets. The depression of the melting temperature with mixing ensures that temperatures along the adiabats in our models are above melting temperatures \citep{Fei2021}.  Similar
arguments apply to Fe mixed with hydrogen, for which dissolved H also
stabilizes the liquid relative to the solid and lowers the liquidus.

For the structure of the envelope we integrate numerically the hydrostatic and mass conservation equations as in the interior (Equations \ref{eq:dmdr} and \ref{eq:dPdr}) using the tabulated function from \cite{Chabrier2019} for the EoS. The temperature structure of the envelope is obtained by numerically integrating the coupled hydrostatic, mass conservation, and energy transport equations outward from the magma
ocean surface, followed by an iterative inward integration from the optically-thin outer atmosphere. Throughout most of the envelope, energy transport is convective, and the temperature gradient is therefore taken to
be the convective (pseudoadiabatic) gradient,

\begin{equation}
\frac{dT}{dr} = -\,\nabla T_{\rm conv},
\end{equation}
where $\nabla T_{\rm conv}$ is evaluated using a moist pseudoadiabat appropriate
for an H$_2$--silicate vapor mixture \citep{Graham_2021}, scaled to the H$_2$
equation of state \citep{Chabrier2019}. Above the convective layer where the gas is still collisional, radiative diffusion applies,

\begin{equation}
\nabla T_{\rm rad}
=
-\frac{3 \kappa \rho L_{\rm int}}{64 \pi \sigma T^3 r^2},
\label{eqn:nablarad}
\end{equation}
where $\kappa$ is the local opacity, $\rho$ the density, and $L_{\rm int}$ the
intrinsic luminosity carried upward from the interior. The envelope transitions smoothly into a radiative regime by relaxing the temperature gradient with Eddington’s
two-stream solution for a plane-parallel grey atmosphere \citep{Guillot2010},

\begin{equation}
\begin{aligned}
T^{4}(\tau)
&=
\frac{3}{4}\,T_{\rm int}^{4}
\left(\tau+\frac{2}{3}\right) \\
+&
\frac{3}{4}\,T_{\rm eq}^{4}
\left[
\frac{2}{3}
+\frac{2}{3\gamma_{\rm vis}}
+
\left(
\frac{\gamma_{\rm vis}}{3}
-\frac{2}{3\gamma_{\rm vis}}
\right)
e^{-\gamma_{\rm vis}\tau}
\right],
\end{aligned}
\label{eqn:Eddington}
\end{equation}
where $\tau$ is the optical depth measured downward from the top of the
atmosphere, $T_{\rm eq}$ is the equilibrium temperature set by stellar
irradiation,  $T_{\rm int}$ is the intrinsic temperature associated with the
interior luminosity, and $\gamma_{\rm vis}$ in this context is the visible-to-thermal opacity ratio. We impose conservation of  intrinsic luminosity, rather than a constant intrinsic flux as assumed in the plane-parallel equations, by accounting for the area dependence of the intrinsic flux $f_{\rm int}(r) = \sigma T_{\rm int}^4(r)$ such that $T_{\rm int}(r) = T_{\rm int, s} (r_s/r)^{1/2}$ where subscript $s$ signifies evaluation just above the supercritical magma ocean.  At large infrared optical depth ($\tau \gg 1$), Equation~\ref{eqn:Eddington} approaches the radiative--diffusion limit, yielding a radiative temperature gradient consistent with Equation~\ref{eqn:nablarad} and thus ensuring consistency with deep-envelope energy transport. This approach affords a smooth and numerically stable transition from the deep convective adiabat to the irradiated upper atmosphere.

\subsection{Luminosity}
The intrinsic luminosities of sub-Neptunes are largely unknown and model dependent.  Most often evolution models specify an initial entropy and intrinsic luminosity, $L_{\rm int}$, based on inferences about retention of heat during accretion. However, this approach does not guarantee self-consistency; a specified $L_{\rm int}$ may exceed, or fall short of, the flux the atmosphere can transmit given its opacity structure and boundary conditions, leading to either interior heating or unphysical cooling timescales.  A closure for this problem, and thus self consistent values for $L_{\rm int}$, can be obtained from consideration of boundary layer effects.

While the bulk of the lower region of the hydrogen-rich envelope is convective, a third layer at the base of the atmosphere arises from a boundary layer between the supercritical melt and the overlying convective region in the envelope \citep{Ahlers_2009}, augmented in many cases by the hindrance of convection due to the mass loading of heavy elements at relatively high temperatures \citep{Misener2023}. The lower boundary layers restrict the intrinsic upward luminosities that comprise heat transfer from the core to the atmosphere, forming a bottleneck. 

The minimum thickness of the basal boundary layer separating the interior (core) and the overlying envelope, $\delta$, comes from consideration of the boundary for Rayleigh-B\'ernard convection in the envelope.  The width of the boundary layer is controlled by the vigor and scale of the convection and is given by scalings similar to 
\begin{equation}
\delta \sim n h\,\mathrm({Ra/Ra_{\rm c}})^{-1/3}, 
\label{eqn:delta}
\end{equation}
where $h$ is a characteristic convecting length scale, taken to be some multiple $n$ of the pressure scale height
evaluated at the base of the atmosphere,  $Ra$ is the Rayleigh number and $Ra_{\rm c}$ is the critical Rayleigh number for the convecting H$_2$-rich 
envelope \citep[e.g.,][]{Long_20202, Grossmann_2000}. The Rayleigh number is computed as
\begin{equation}
Ra = \frac{g\,\alpha\,\Delta T\,h^3}{\nu\,\kappa},
\label{eq:ra}
\end{equation}
where $h$ is here taken to be $5H$ for scaleheight $H$, $\Delta T$ is the temperature contrast across the layer, $\nu$ is the kinematic viscosity, $\kappa$ is the thermal diffusivity, and $\alpha$ is the thermal expansivity. Use of the adiabat for $\Delta T$, rather than superadiabatic temperatures, likely overestimates $Ra$, resulting in underestimates for the widths of the boundary layers (see below).  Kinematic viscosities for H$_2$ are estimated here using a Sutherland-law fit for dynamic viscosities $\eta$: 

\begin{equation}
\eta(T) = \eta_0 \left(\frac{T}{T_0}\right)^{3/2}\frac{T_0 + S_{T}}{T + S_{T}},
\label{eqn:dynamic_viscosity}
\end{equation}
with reference viscosity $\eta_0$ at reference temperature $T_0$ and Sutherland temperature $S_T$ for H$_2$ from \cite{Braun2018}.  The kinematic viscosities are then 

\begin{equation}
\nu = \frac{\eta}{\rho}.
\label{eqn:kinematic_viscosity}
\end{equation}
Thermal expansivities required to evaluate $Ra$ are obtained using 
\begin{equation}
\alpha = -\frac{1}{T}\left(\frac{\partial\log\rho}{\partial\log T}\right)_{\!P}
\label{eqn:alpha}
\end{equation}
based on the EoS tables for hydrogen from \cite{Chabrier2019}.  We adopt a thermal diffusivity $\kappa$  of  $1.0\times10^{-5}$ m$^2$/s for the purpose of evaluating approximate values for the Rayleigh number. For nominal parameters relevant for sub-Neptunes in this study, e.g.,  $\Delta T = 3000$ K, surface pressure $P_s = 1$ GPa, planet mass $M_p=6M_{\oplus}$, and density $\rho = 47$ kg/m$^3$,  one obtains $Ra \simeq 10^{32}$. 

The critical Rayleigh number depends on rotation rate, among other factors \citep{Aurnou2020}, through the Ekman number ($Ek = \nu/(2\Omega h^2)$ where $\nu$ is the kinematic viscosity, $\Omega$ is the rotation period of the body, and $h$ is the relevant height. A commonly used scaling in the limit where $Ek \rightarrow 0$ is \citep{Chandrasekhar1961}:

\begin{equation}
Ra_{\rm c} = 8.7 Ek^{-4/3}.
\label{eqn:Ekman}
\end{equation}
The Ekman number for a $6M_{\oplus}$ sub-Neptune with the atmosphere used to evaluate $Ra$ described above, where $\nu = 9.6\times 10^{-7}$ m$^2$/s and $H \sim 10^5$ m, and tidally-locked rotation periods from 10 days to 100 days, is $2\times 10^{-15}$ to $2\times 10^{-14}$, yielding $R_{\rm c}$ values of $3\times 10^{20}$ to $2\times 10^{19}$, respectively. When paired with the corresponding Rayleigh numbers of $2\times 10^{32}$, $Ra/Ra_{\rm c}$ ranges from $10^{11}$ to $10^{13}$ and the boundary layer thicknesses implied by Equation \ref{eqn:delta} are of order 20 to 100 meters.  We can combine these thicknesses with opacities to derive the optical depth for the radiative boundary layer, $\tau_{\rm BL}=\kappa \rho \delta$.   

The intrinsic luminosity $L_{\rm int}$ is determined from the blackbody emission $L_{\rm bb} = 4\pi r_{\rm s}^2 \sigma T_{\rm s}^4$ using the the ability of the magma ocean to radiate through the boundary layer immediately above its surface, which for a grey atmosphere is,

\begin{equation}
L_{\mathrm{int}} = L_{\rm bb}
\left(
\frac{4}{3(\tau_{\rm BL}+1)}
\right).
\label{eqn:Lint}
\end{equation}

\noindent The intrinsic temperature at the surface satisfies 

\begin{equation}
L_{\rm int} = 4 \pi r_{\rm s}^2 \sigma T_{\rm int, s}^4.
\label{eqn:Tint}
\end{equation}
This formulation allows for self consistency between the surface of the magma ocean and the upper atmosphere, accommodating the link between intrinsic luminosity in the regime of vigorous convection. Rather than resolving the temperature gradient in this often very narrow boundary, its influence as a governor on radiative flux is captured through Equation \ref{eqn:Lint}

The basal radiative layer can be further stabilized and enhanced in elevation by the molecular weight gradients that attend the interaction between the supercritical magma ocean and the overlying envelope.  As a result, the boundary between the surface of the magma ocean and the convective layer in the envelope can be both pronounced and explicitly resolvable in our models. Rather than allowing arbitrarily large intrinsic luminosities that would dynamically destroy stratification, at least temporarily, we adopt a quasi-static closure in which a Ledoux-stable layer, when permitted thermodynamically, forms, persists and therefore acts as a finite thermal resistance that limits the transmitted flux. Thermal bottlenecks created by compositional gradients in the interiors of sub-Neptunes were described by \cite{Vazan2018} and \cite{arevalo_sub-neptune_2025}.

Stability against convection in the presence of composition gradients is governed by the Ledoux criterion:
\begin{equation}
\nabla_{\rm rad} > \nabla_{\rm ad} + \frac{\phi}{\delta}\,\nabla_\mu ,
\label{eqn:Ledoux_criterion}
\end{equation}
where
\begin{align}
\nabla_\mu &\equiv \frac{d\ln \mu}{d\ln P}, \\[6pt]
\delta &\equiv -\left(\frac{\partial \ln \rho}{\partial \ln T}\right)_{P,\mu}, \\[6pt]
\intertext{and}
\phi &\equiv \left(\frac{\partial \ln \rho}{\partial \ln \mu}\right)_{P,T}.
\label{eqn:Ledoux}
\end{align}
Here $\mu$ is the mean molecular weight. A positive molecular-weight gradient, ($\nabla_\mu > 0$), increases the critical temperature gradient required for convective instability by the factor $(\phi/\delta)\nabla_\mu$. The ratio $(\phi/\delta)$ is unity for an ideal gas, and remains of order unity for reasonable values for compressibility factor $Z$, so in practice we simplify the Ledoux criterion (Equation \ref{eqn:Ledoux_criterion}) to $\nabla_{\rm rad} > \nabla_{\rm ad} + \nabla_\mu$.  Where the Ledoux-stable region spans a finite radial extent, it further limits $L_{\rm int}$ for the overlying atmosphere layers. The intrinsic luminosity  is limited by the requirement that the layer remain marginally Ledoux-stable. In this radiative layer, the upward energy flux is given by the diffusion approximation,
\begin{equation}
F_{\rm rad}
=
-\frac{16 \sigma T^{3}}{3 \kappa \rho}\,\frac{dT}{dr},
\end{equation}
which may be written in terms of dimensionless gradient $\nabla$ as
\begin{equation}
F_{\rm rad}
=
\frac{4 \sigma T^{4}}{3 \kappa \rho H}\,\nabla ,
\end{equation}
where $H=P/(\rho g)$ is the pressure scale height and $\nabla \equiv d\ln T/d\ln P$. The maximum radiative flux consistent with stability against convection is obtained by evaluating this expression at the largest temperature gradient permitted by the Ledoux criterion:
\begin{equation}
\nabla_{\max}
=
\nabla_{\rm ad}
+
\frac{\phi}{\delta}\,\nabla_{\mu}.
\end{equation}

In the radiative layer, we therefore have,
\begin{equation}
\nabla_{\rm rad}= \nabla_{\rm max} =
\frac{3 \kappa P L}{64 \pi \sigma G M T^{4}},
\label{eqn:nabla_rad}
\end{equation}
leading to a definition for the the maximum luminosity that can be transported without violating
Ledoux stability:
\begin{equation}
L_{\mathrm{Ledoux}}
=
\left(
\frac{64 \pi \sigma G M T^{4}}{3 \kappa P}
\right)
\nabla_{\rm max}
\label{eqn:L_ledoux}
\end{equation}
evaluated at a representative location near the top of the Ledoux-stable region. The intrinsic luminosity is then capped such that
$L_{\rm int} = \min(L_{\rm int}',\,L_{\mathrm{Ledoux}})$, where $L_{\rm int}'$ is the value
in the absence of the stable boundary. This upper limit is applied iteratively during the solution
procedure until a self-consistent thermal structure satisfying Ledoux stability is obtained.
Any larger luminosity would require a radiative temperature gradient steeper than $\nabla_{\rm max}$, causing $\nabla_{\rm rad}$ to exceed the Ledoux stability limit. In that case the stabilizing molecular-weight gradient would no longer suppress convection, and the layer would become convectively unstable.

Expressing the radiative flux as a finite difference across the
Ledoux-stable region, we have
\begin{equation}
L_{\rm Ledoux}
\simeq
4\pi r_{\nabla_{\mu,\rm max}}^{2}\,
\frac{4\sigma}{3}\,
\frac{T_{\rm lower}^{4}-T_{\rm upper}^{4}}{\Delta\tau_{\rm Ledoux}},
\label{eqn:L_ledoux_tau}
\end{equation}
which makes explicit that the transmissible luminosity is controlled by the
integrated optical-depth thickness across the inhibited region $\Delta\tau_{\rm Ledoux}$.
Here $T_{\rm lower}$ and $T_{\rm upper}$ are the temperatures at the lower and upper
boundaries of the layer, and $r_{\nabla_{\mu, \rm max}}$ is a representative radius within the
region where $\nabla \mu$ is maximized. Equation~(\ref{eqn:L_ledoux_tau}) therefore represents the integrated radiative
throughput of the Ledoux-stable layer and provides a global maximum for the intrinsic
luminosity. It also avoids numerical oscillations that can arise from high-frequency layer-by-layer
adjustments of the temperature gradient.

Molecular weight gradients specify where
stratification is stable against overturn. In the presence of a phase change, an additional threshold must obtain for stability due to the thermal effects of the enthalpy of reaction. Accordingly, we include a microphysical inhibition threshold, following \cite{Markham2022}, based on a critical ``heavy species'' mole-fraction threshold $x_{\rm inhib}$,
\begin{equation}
\begin{aligned}
x_{\rm inhib}
&=
\left[
\left(\frac{\mu_{\rm melt} \Delta H}{R T}-1\right)
(\epsilon-1)
\right]^{-1},
\end{aligned}
\label{eqn:x_inhib}
\end{equation}
where  $\mu_{\rm melt}$ is the mean molecular
weight of the condensed phase, $\epsilon \equiv \mu_{\rm vap} / \mu_{\rm H_2}$ where  $\mu_{\rm vap}$ is the mean molecular
weight of the heavy condensing vapor species (e.g., Mg, SiO), and $\Delta H$ is the effective latent heat per unit mass for condensation. Physically, $x_{\rm inhib}$ encodes stable stratification in the presence of the phase change. In practice, the Ledoux-stable, inhibited region is activated only when both a stabilizing molecular-weight gradient exists ($\nabla_\mu > 0$)  and
the heavy-component abundance exceeds the compositional threshold
($x_{\rm heavy} > x_{\rm inhib}$). We combine these criteria into a smooth inhibition ``strength" $w \in [0,1]$ that controls
how strongly the local temperature gradient is driven toward the radiative (diffusion-
limited) gradient,
\begin{equation}
\begin{aligned}
w
&= w_\mu\, w_{\rm x_{inhib}},
\\[0.5ex]
w_\mu
&= \mathcal{H}\!\left(\nabla_\mu\right),
\\[0.5ex]
w_{\rm x_{inhib}}
&= \mathcal{H}\!\left(\frac{x_{\rm heavy}}{x_{\rm inhib}}\right),
\end{aligned}
\label{eqn:w_strength}
\end{equation}
where $\mathcal{H}$ is a Hill function used to smooth the transition into and out of the
inhibited regime and to avoid numerical noise.

The temperature gradient is evaluated by interpolating between a baseline gradient
and the radiative-required gradient,
\begin{equation}
\nabla
=
(1-w)\,\nabla
+
w\,\nabla_{\rm rad},
\end{equation}
where $\nabla=\nabla_{\rm ad}$ in convective regions and
$\nabla=\nabla_{\rm rad}$ otherwise. In the limit $w\rightarrow 1$, the
temperature gradient approaches the diffusion-limited value, while for $w\rightarrow 0$
the solution reverts to the transport regime outside the Ledoux-stability region.

Both types of basal boundary layers are global bottlenecks for the intrinsic luminosity. 
When a Ledoux-stable compositional gradient
is present,  $L_{\rm int}$ is capped at the maximum luminosity that can be transmitted through
the inhibited layer without violating Ledoux stability. This constraint is conceptually
analogous to the luminosity limit imposed by the basal thermal boundary layer of thickness
$\delta_{\rm BL}$, which bounds the net heat flux delivered from the magma ocean into the
envelope. The distinction between the two lies in resolvability: the Ledoux-stable region
is sufficiently extended that its stabilizing gradient can be described explicitly, whereas
the basal thermal boundary layer is often too thin to resolve and is therefore parameterized as a
global flux limit.

We treat heat transport in the non-conective hydrogen boundary layer as radiative, neglecting conduction. To illustrate why, we
compare an effective radiative conductivity $\lambda_{\mathrm{rad}} = 16\,\sigma T^{3}/(3\,\kappa\,\rho)$ with the thermal conductivity
$\lambda_{\mathrm{cond}}$ derived from the non-electronic, molecular  $\lambda_{\mathrm{cond}}$ of $ \sim 2\,\mathrm{W\,m^{-1}\,K^{-1}}$ appropriate for our model conditions  \citep{French2012}.  At the conditions that obtain for a typical stable boundary layer ($T \sim 3000\,\mathrm{K}$, $P \sim 5\,\mathrm{GPa}$),
$\kappa \approx 20\,\mathrm{cm^{2}\,g^{-1}}$ \citep{freedman2014a}, yielding 
$\lambda_{\mathrm{rad}} \approx 14\,\mathrm{W\,m^{-1}\,K^{-1}}$, exceeding
$\lambda_{\mathrm{cond}}$ $\sim\!2\,\mathrm{W\,m^{-1}\,K^{-1}}$ by a
factor of seven. Because of the marked temperature
dependence of the mean opacity whereby $\kappa$ rises by more than an order of
magnitude between $3000$ and $5000\,\mathrm{K}$, faster than $T^{3}$, there is a radiative/conductive crossover near
$4300$--$4900\,\mathrm{K}$ depending on density, above the temperatures in our boundary layers. Accordingly, our result is 
consistent with \citet{Misener2023}, who assumed higher temperatures of $\sim\!5000\,\mathrm{K}$ at the base of the atmosphere, where
 $\kappa \gtrsim 600\,\mathrm{cm^{2}\,g^{-1}}$, resulting in conduction competing with radiative diffusion. The difference reflects  the $\sim\!2000\,\mathrm{K}$ difference in base temperature.

\subsection{Planet Radius}
The planetary radius for these models is defined using the chord optical depth \citep{Guillot2010}, coupled with opacities from \cite{freedman2014a} evaluated at metallicities corresponding to the local gas composition. The binodal interface between the magma ocean and the overlying envelope constrains the surface temperature at a given pressure to a relatively narrow range, such that the thermal state of the system is expressed primarily through the radial position of the boundary rather than through large variations in temperature.  Since the atmospheric entropy and scale height are controlled by the intrinsic luminosity, the basal boundary layer directly governs the achievable planetary radius. In this sense, the core–envelope boundary layers provide upper bounds on the radii of sub-Neptunes.

\subsection{Physical Chemistry}
As a result of retention of significant fractions of gravitational potential energy during accretion, sub-Neptunes are expected to have formed hot enough that a global magma ocean was present beneath their primary hydrogen-rich atmospheres \citep[e.g.,][]{Young_2024}. Cooling times moderated by the dense atmospheres are such that the magma oceans have likely persisted over Gyr timescales, meaning that a significant fraction of sub-Neptunes observed today should have magma oceans beneath their envelopes  \citep{ginzburg2016a, Misener2022}. Therefore, to understand the chemistry and structure of sub-Neptunes requires an understanding of the interaction between the hydrogen-rich envelopes and melts, or supercritical fluids, comprising the cores of the planets \citep{Kite_2020b, Schlichting_Young_2022}. 

Recent work has applied advances in our understanding of the physical chemistry of H$_2$-silicate-Fe metal systems to sub-Neptunes \citep{Young_2024, gilmore_core-envelope_2025, Young2025_Differentiation, RogersYoungSchlichting2025_MNRAS}. That work shows that the interface separating  the outer envelope of a sub-Neptune (i.e., its atmosphere, including the supercritical fluid phase where appropriate) and the underlying magma ocean is a phase boundary that can be well approximated as the binodal (or solvus) surface in the MgSiO$_3$-H$_2$ system \citep{gilmore_core-envelope_2025, Young_2024, RogersYoungSchlichting2025_MNRAS}. This boundary moves inward towards the center as the planet cools \citep{RogersYoungSchlichting2025_MNRAS}.  Interior to the boundary, the magma ocean is a supercritical mixture of silicate (modeled as MgSiO$_3$) and hydrogen with or without a distinct Fe-rich metal phase.  For planets with sufficient hydrogen that their cores  have greater than about $1\%$ by mass H$_2$, Fe metal, MgSiO$_3$ (silicate) and H$_2$ are entirely miscible \citep{Young2025_Differentiation}.  In these cases, the cores are a single phase throughout except at the most shallow depths. We emphasize that the binodal phase boundary replaces the notion of a melt–envelope interface imposed at a specified surface. In that picture, the pressure and temperature at the interface are set by the overlying envelope mass, gravity, and the envelope adiabat. In our model, by contrast, the phase diagram fixes the pressure–temperature condition for coexistence, and hydrostatic equilibrium determines the radius at which the phase change occurs in the planet.

\begin{figure*}[t]
\centering
\begin{minipage}[t]{0.47\textwidth}
    \centering
    \includegraphics[width=\textwidth]{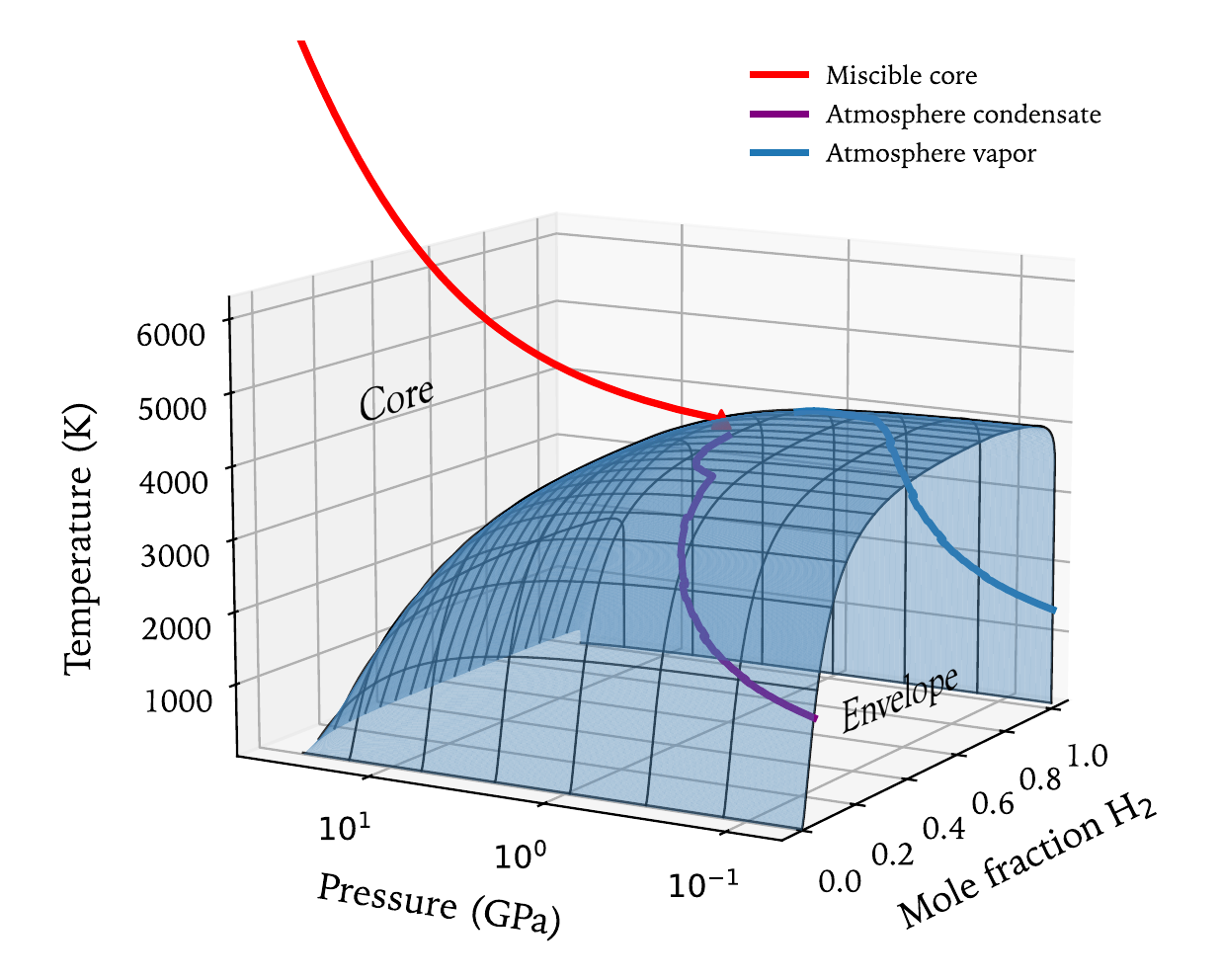}
    \caption{The binodal surface separating the core and envelope for a 
    6~$M_{\oplus}$, 3~wt\,\% H$_2$ sub-Neptune with a fully miscible core 
    prior to Gyr timescales of cooling. The paths through the one-phase core 
    and through the two-phase envelope are shown.}
    \label{fig:3D_early}
\end{minipage}
\hfill
\begin{minipage}[t]{0.47\textwidth}
    \centering
    \includegraphics[width=\textwidth]{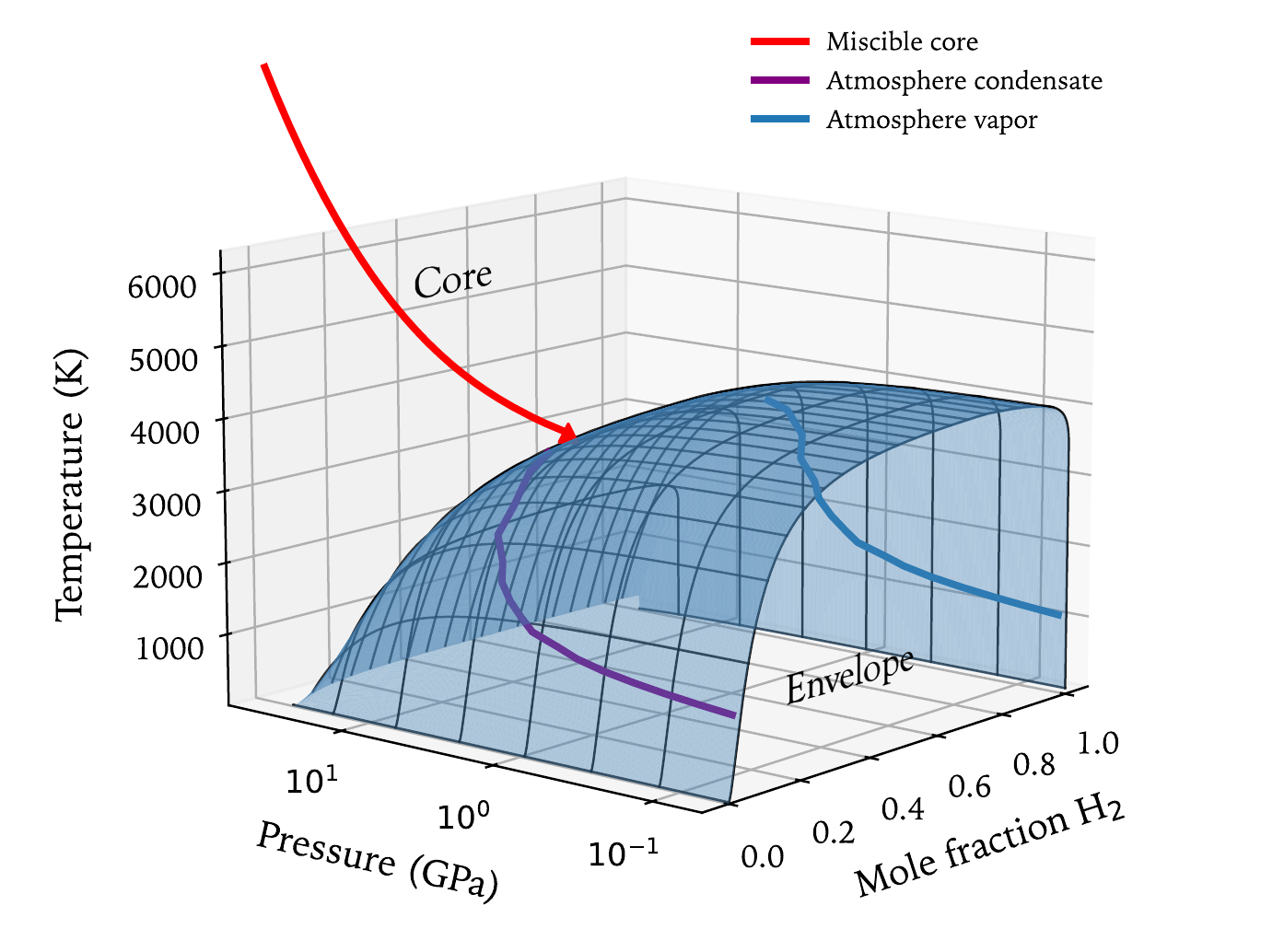}
    \caption{The binodal surface separating the core and envelope for a 
    6~$M_{\oplus}$, 3~wt\,\% H$_2$ sub-Neptune with a fully miscible core 
    after approximately a Gyr time interval of cooling. The paths through 
    the one-phase core and through the two-phase envelope are shown.}
    \label{fig:3D_late}
\end{minipage}
\end{figure*}

\subsubsection{Core-Envelope Boundary}
By determining the common tangents to the Gibbs free energy of mixing in the MgSiO$_3$-H$_2$ system, using the convex hull for the function for example, the values for the mole fractions of hydrogen, $x_{\rm H_2}$, for melt and vapor in equilibrium are obtained.  These coexisting compositions at various $P$ and $T$ values define the binodal surface that separtes cores from overlying envelopes.
 The non-ideal mixing between MgSiO$_3$ and H$_2$ calculated by \cite{gilmore_core-envelope_2025} is used here. Their density functional theory molecular dynamics (DFT-MD) simulations yield the binary free energy of mixing expression  

 \begin{equation}
\begin{split}
\Delta \hat{G}_{\text{binary mix}} &=
\left(L_{\mathrm{CB}} x_{\rm H_2}
+ L_{\mathrm{BC}} (1-x_{\rm H_2})\right)
x_{\rm H_2} (1-x_{\rm H_2}) \\
&\quad \times
\left( 1 - \frac{T}{\tau} + \frac{P}{\pi} \right) \\
&\hspace{-2.0em} + RT \left(
x_{\rm H_2} \ln x_{\rm H_2}
+ (1-x_{\rm H_2}) \ln (1-x_{\rm H_2})
\right),
\end{split}
\label{eqn:MgSiO3_H2}
\end{equation}
where $L_{\mathrm{CB}}$ and $L_{\mathrm{BC}}$ are the binary interaction parameters and $x_{\rm H_2}$ is the mole fraction of H$_2$ along the binary join. The values for the subregular mixing parameters that fit  DFT-MD simulations are:  $L_{\mathrm{CB}} = 622000$ J/mol,  $L_{\mathrm{BC}} = -4950$ J/mol, $\tau = 4800$ K,  and $\pi = -35$ GPa.  
\cite{gilmore_core-envelope_2025} find that addition of minor elements like Al has minimal effects on the position of the binodal surface in $P$-$T$-$x_{H_2}$ space.

Using reasonable estimates for initial specific entropies and associated luminosities, \cite{RogersYoungSchlichting2025_MNRAS} showed that the boundary between the core and envelope for younger sub-Neptunes (age of $10^7$ yr) is predicted to be near 1 GPa and 4000 K for a 6 $M_{\oplus}$ sub-Neptune with 3 weight percent total hydrogen.  With cooling over a Gyr, the boundary shifts inward and is characterized by higher pressures and lower temperatures in the vicinity of 4 GPa and 3000 K.  The specific conditions at the binodal surface depend on the bulk H$_2$ composition of the planet as well as its mass and equilibrium temperature. For comparison, and using the topology from \cite{Young_2024} and \cite{RogersYoungSchlichting2025_MNRAS}, the binodal surfaces for the example 6 $M_{\oplus}$, 3$\%$ H$_2$ fully-miscible planet for the early and late cases based on the current modeling are shown in Figures \ref{fig:3D_early} and \ref{fig:3D_late}. 

Note that the piercing point of the path through the core (red curve) that defines the magma ocean-envelope interface occurs at higher $P$ and lower $T$ for the older planet in Figure \ref{fig:3D_late}. Calculations such as these illustrate that the pressure of the binodal is a surrogate for the relative age of a sub-Neptune, all else equal. The binodal separating the core from the envelope spans a significantly narrower range of conditions  than the base of atmospheres in ``classical" models where there is no interaction between hydrogen in the envelope and the magma ocean core \citep{Young_2024, RogersYoungSchlichting2025_MNRAS}.

\subsubsection{Core Miscibility}
\label{section:miscibility}
In order to determine the state of the core given the planet's pressure-temperature structure and its  mass fraction of hydrogen we make use of the ternary phase equilibria from \cite{Young2025_Differentiation}.  With this approach, we construct a thermodynamically consistent ternary mixing model for miscible
MgSiO$_3$–Fe–H$_2$ melts by extrapolating three independently constrained binary
systems (MgSiO$_3$–H$_2$, MgSiO$_3$–Fe, and Fe–H$_2$) into ternary composition space.
Phase stability is determined by minimizing the molar Gibbs free energy of mixing
at fixed pressure and temperature. The total free
energy of mixing is written as the sum of an ideal configurational entropy term
and a non-ideal (excess) mixing term. The ideal contribution is given by
$RT\sum_i x_i \ln x_i$, where $x_i$ are the mole fractions of Fe (A), H$_2$ (B),
and MgSiO$_3$ (C). Non-ideal interactions are parameterized using a subregular
solution formalism, with interaction parameters taken directly from published
DFT-MD simulations and experimental constraints on the three binary joins.

The resulting Gibbs free energy of mixing for the ternary system is obtained following the Muggianu–Jacob projection method:

\begin{equation}
\begin{aligned}
&\Delta \hat{G}_{\text{mix}} = RT\!\left(
x_{\mathrm{A}}\ln x_{\mathrm{A}} +
x_{\mathrm{B}}\ln x_{\mathrm{B}} +
x_{\mathrm{C}}\ln x_{\mathrm{C}}
\right) \\
& + (1/2)x_{\mathrm{A}} x_{\mathrm{B}} \left( L_{\mathrm{AB}} (1 + x_{\mathrm{B}} - x_{\mathrm{A}}) + L_{\mathrm{BA}} (1 + x_{\mathrm{A}} - x_{\mathrm{B}}) \right) \\
& + (1/2) x_{\mathrm{B}} x_{\mathrm{C}} \left( L_{\mathrm{BC}} (1 + x_{\mathrm{C}} - x_{\mathrm{B}}) + L_{\mathrm{CB}}  (1 + x_{\mathrm{B}} - x_{\mathrm{C}}) \right) \\
& (1 - T/\tau + P/\pi) \\
& + (1/2) x_{\mathrm{C}} x_{\mathrm{A}} \left( L_{\mathrm{CA}}  (1 + x_{\mathrm{A}} - x_{\mathrm{C}}) + L_{\mathrm{AC}}  (1 + x_{\mathrm{C}} - x_{\mathrm{A}}) \right)
\end{aligned}
\label{eqn:Gmix}
\end{equation}

\noindent where $x_{\mathrm{A}}$, $x_{\mathrm{B}}$, and $x_{\mathrm{C}}$ denote the mole
fractions of Fe, H$_2$, and MgSiO$_3$, respectively, with
$x_{\mathrm{A}}+x_{\mathrm{B}}+x_{\mathrm{C}}=1$.
The MgSiO$_3$–H$_2$ interaction parameters are
$L_{\mathrm{CB}}=6.22\times10^{5}$~J~mol$^{-1}$ and
$L_{\mathrm{BC}}=-4.95\times10^{3}$~J~mol$^{-1}$, with temperature and pressure
dependence governed by $\tau=4800$~K and $\pi=-35$~GPa, as fitted by
\citet{gilmore_core-envelope_2025}.
The MgSiO$_3$–Fe interaction is treated as a regular solution with
$L_{\mathrm{AC}}=L_{\mathrm{CA}}=2.4\times10^{5}-28T+1116P$ (J~mol$^{-1}$), based on
DFT-MD calculations for MgO–Fe mixing.
The Fe–H$_2$ interaction parameters are
$L_{\mathrm{AB}}=1.38\times10^{5}-9500P$ and
$L_{\mathrm{BA}}=1.7\times10^{4}-9500P$ (J~mol$^{-1}$), calibrated to reproduce
experimental and \emph{ab initio} constraints on hydrogen solubility in molten iron.
No explicit, and unknown, ternary interaction
parameter is included ($L_{\mathrm{ABC}}=0$) as is often the case in such applications.

With this free energy of mixing we can map the regions of miscibility and immiscbility between MgSiO$_3$, Fe, and H$_2$ as a function of pressure and temperature (a python code for these calculations, \texttt{G\_planet\_intersection}, can be found at \url{https://github.com/eyoungucla/G_plane_intersection}).

\subsubsection{Material Properties of the Core}
\label{section:Material_properties}
For liquid silicate densities, we use the approach from \cite{Young2025_Differentiation}.  Briefly, for the silicate melt we use an equation of state  fit to the MgSiO$_3$ liquid properties determined by \cite{DeKoker2009} using the algorithms of \cite{Wolf_2018}. Total pressure is computed by combining the elastic (cold) compression term from a Vinet equation of state with thermal pressure contributions, following the formulation of \citet{Wolf_2018}.
 For liquid iron, we use a Vinet equation of state modified to include thermal pressure from  \cite{Kuwayama2020}. The Gr\"uneisen parameter is a simple function of the compression ratio, which in turn determines the thermal contribution to pressure.

In our models here, we assume  that addition H and, in some cases Fe, to the supercritical magma ocean changes its density, but does not significantly affect bulk moduli and their pressure derivatives (we investigate this assumption in future publications).  
For regions of the planet inside the binodal, hydrogen and  MgSiO$_3$ melt are completely miscible. Based on DFT-MD simulations \citep{Young2025_Differentiation}, the presence of hydrogen in the silicate melt reduces its density such that a linear mixture of the compressed densities of MgSiO$_3$ and H$_2$ by volume reproduces the density of the mixture at the conditions of the binodal.  We therefore include the effects of H$_2$ on the density of the supercritical melt by calculating the density of the mixture at the pressures and temperatures of the surface of the magma ocean, $\rho_0$. The density of the mixture is  
\begin{equation}
    \frac{1}{\hat{V}_{\rm mix}} = x_{\text{H}_2}\frac{1}{\hat{v}_{\text{H}_2}} + x_{\text{sil}} \frac{1}{ \hat{v}_{\text{sil}}},
\label{eq:Vmix}
\end{equation}
where $\hat{v}_{\text{H}_2}$ and $\hat{v}_{\text{sil}}$ are the molar volumes, and where $x_{\text{H}_2}$ and $x_{\text{sil}}$ are the mole fractions of hydrogen and silicate in the binary mixture. Densities are obtained from MW$_{\rm mix}/\hat{V}{\rm mix}$ where MW$_{\rm mix}$ is the molecular weight of the mixture. Molar volumes are obtained from densities and the equations of state using $\rho_i/{\rm MW}_i$.  We fixed the densities to be $0.09$ g $\rm cm^{-3}$ and $2.5$ g $\rm cm^{-3}$ for H$_2$ and silicate, respectively, at the surface of the magma ocean. These values come from the equations of state for hydrogen from \cite{Chabrier2019} and {\it ab initio} molecular dynamics simulations of MgSiO$_3$ melt at 6000 K and 3.5 GPa \citep{Young2025_Differentiation}.   

Following \cite{Young2025_Differentiation}, we model Fe dissolved in a supercritical silicate--H$_2$ melt by assigning Fe a partial molar volume that reflects Fe occupation of Mg-like cation sites within the silicate melt framework. The Fe-site  partial molar volume is
\begin{equation}
\bar V_{\mathrm{Fe\, site}}^{\mathrm{eff}}
= \phi\,\alpha\;\bar V_{\mathrm{MgSiO_3}}^{\mathrm{pure}}(P,T),
\end{equation}
and the corresponding effective partial Fe density is
\begin{equation}
\rho_{\mathrm{Fe\, site}}^{\mathrm{eff}}= \frac{{\rm MW}_{\mathrm{Fe}}}{\bar V_{\mathrm{Fe\, site}}^{\mathrm{eff}}},
\end{equation}
where $\phi$ denotes the fraction of the pure MgSiO$_3$ formula-unit molar volume attributable to the Mg site,  $\alpha=(r_{\mathrm{Fe}}/r_{\mathrm{Mg}})^3$ scales the site volume by the ratio of octahedrally-coordinated ionic radii \citep{Shannon_1976}, and $\bar V_{\mathrm{MgSiO_3}}^{\mathrm{pure}}={\rm MW}_{\mathrm{MgSiO_3}}/\rho_{\mathrm{MgSiO_3}}^{\mathrm{pure}}$ is the molar volume of pure MgSiO$_3$ at the local $P$ and $T$.
Assuming additivity of specific volumes (no excess mixing volume), the density of the miscible melt with Fe mass fraction $w_{\mathrm{Fe}}$ is given by 
\begin{equation}
\frac{1}{\rho_{\mathrm{mix}}}
= \frac{1-w_{\mathrm{Fe}}}{\rho_{\mathrm{MgSiO_3+H_2}}}
+ \frac{w_{\mathrm{Fe}}}{\rho_{\mathrm{Fe}}^{\mathrm{eff}}},
\end{equation}
where $\rho_{\mathrm{MgSiO_3+H_2}}$ is the density of the MgSiO$_3$+H$_2$ mixture derived at $P$ and $T$. An empirical value for $\phi$ of 0.5 is obtained from the  MgO-Fe system \citep{Insixiengmay_2025}.

\subsection{Evolution with Time}
\begin{figure*}[t]
\centering
   \includegraphics[width=0.85\textwidth]{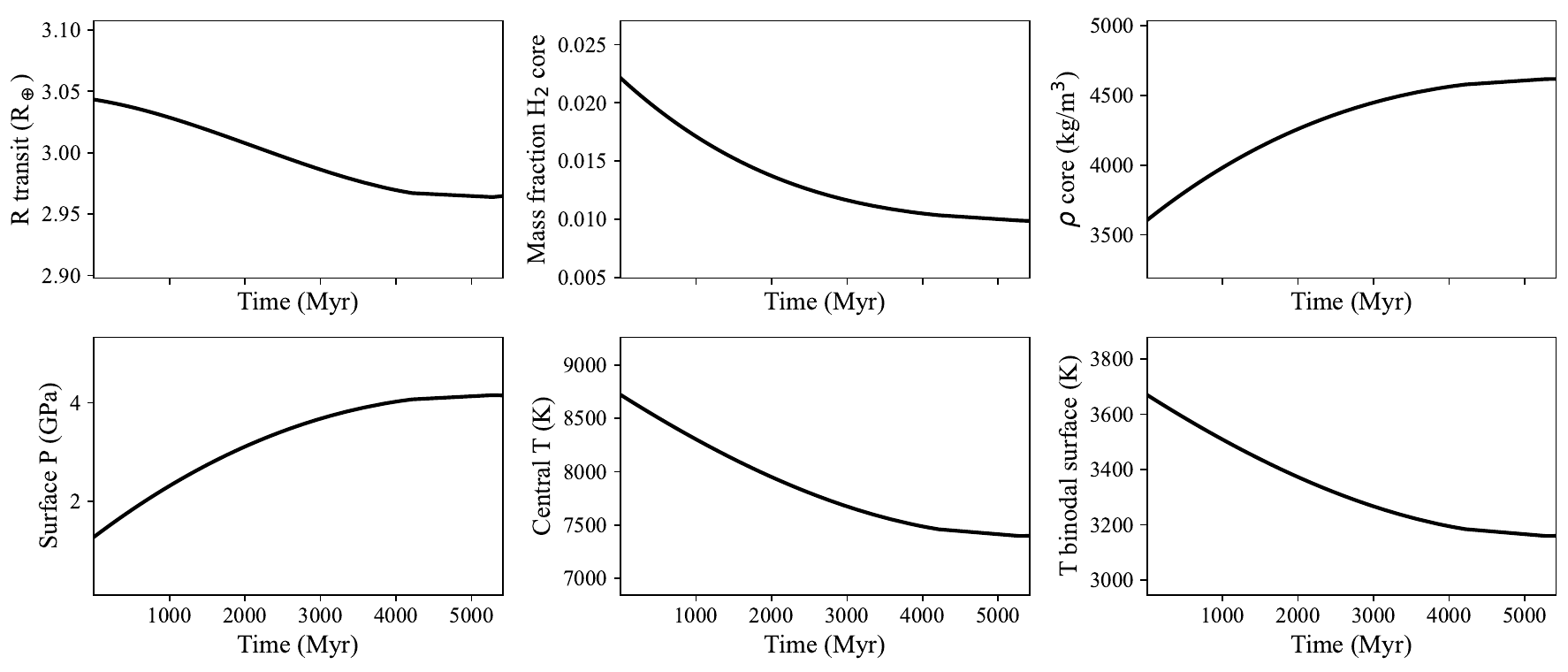}
    \caption{Time evolution of model a sub-Neptune with total mass of 6$M_\oplus$ and mass fractions of H$_2$ of 3\%.  In this example, $Ra/Ra_{\rm c} =\times 10^{12}$, resulting in an initial intrinsic luminosity of $6.3\times 10^{15}$ W that decreases over the approximately 5 Gyr of evolution to $1.1\times 10^{15}$ W.}
\label{fig:Evolution}
\end{figure*}

We do not use determinative evolutionary paths in this population study due to their computational cost, as described above. Instead, we use the pressures of the binodal interface between the sub-Neptune supercritical magma oceans and envelopes as a means of specifying the relative ages of our model planets. Luminosity could also be used \citep{RogersYoungSchlichting2025_MNRAS}, but initial luminosities are not a free parameter in our inverse statistical approach. An example evolutionary path is shown in Figure \ref{fig:Evolution}.  The quasi-statical evolution curves are calculated by specifying the planet mass $M_p$, bulk hydrogen mass fraction,  equilibrium temperature $T_{\rm eq}$, and an initial binodal pressure of $0.9$~GPa. The binodal pressure was increased incrementally. For each step we calculated the total energy of the planet $E_p$ and the associated intrinsic luminosity $L_{\rm int}$. The timestep associated with a step from $P_i$ to $P_{i+1}$ was obtained from energy balance,

\begin{equation}
dt = \frac{\left|E_p(P_{r_{i+1}})-E_p(P_{r_i})\right|}{L_{\rm int}},
\end{equation}
where $L_{\rm int}$ comes from Equation \ref{eqn:Lint}. Total energy $E_p$ is calculated from

\begin{equation}
E_{p} = \int_{0}^{R_{p}}
\left(-\frac{G M(r)}{r} + c\,T(r)\right)\,\rho(r)\,4\pi r^2\,dr,
\label{eqn:E_total}
\end{equation}
where $c$ is the specific heat capacity for the melt or envelope using the equations of state and species described above. Pressure steps are sufficiently small to ensure that $dt$ is approximately three to four orders of magnitude smaller than the cooling time constant  $\tau_{\rm cool} = |E_p|/L_{\rm int}$.

The example evolution track shown in Figure \ref{fig:Evolution} illustrates several salient features relevant to this study. Most importantly, the shifts in radius are relatively small over the timescales of interest defined by the ages of host stars for sub-Neptunes derived from the planet archive (Appendix A).  These ages define a broad log-normal-like distribution with a mode near 2.5 Gyr and a median of 4.3 Gyr. With cooling, the planet's silicate-hydrogen binodal surface moves inward with time \citep{RogersYoungSchlichting2025_MNRAS}. As the binodal moves inward, the pressure at the surface increases from $< 2$ GPa to about 4 GPa and thereafter changes extremely slowly. The timescale for evolution to nearly steady state is of order a few Gyr where luminosity is governed by the the basal radiative boundary, or hundreds of millions of years if this layer were not to be present. The shape of the binodal phase boundary that separates the supercritical magma ocean from the overlying hydrogen-rich envelope dictates the finite range of surface pressures. At lower pressures, the temperature of the interface is higher, but the critical temperature (at the crest of the binodal, the maximum for a given pressure) approaches a maximum of 4350 K at 0 GPa (about 100 K higher than at 1 GPa). Conversely, as pressure exceeds 4 GPa, and the binodal temperature approaches 2000 K for the hydrogen mass fractions of interest here (a few percent or less by mass), no hydrogen would remain in the supercritical magma ocean, and the binodal ceases to be relevant. The evolution curves also illustrate that hydrogen is continually exsolved from the supercritical magma ocean into the envelope, and this process slows considerably with time, as shown previously by \cite{RogersYoungSchlichting2025_MNRAS}. Less massive envelopes (lower initial pressures at the binodal interface) result in greater intrinsic luminosities, and faster evolution than shown in this figure.

\subsection{Atmospheric escape model}
\label{sec:escape}
\subsubsection{Energy-limited Escape}
For each model planet, the likelihood that the envelope (atmosphere) survived for the time interval corresponding to the planet's assigned age is evaluated in the population study.  This is the probabilistic approach to determining which planets have extant envelopes and which should be relegated to stripped cores.  We use an energy-limited prescription for atmospheric loss by irradiation treated as a continuous mass-loss process. The time-integrated mass loss determines whether an atmosphere is retained or stripped. We augmented this by a simple criterion for hydrodynamic escape. 

Atmospheric loss is modeled using an energy-limited prescription based on the atmospheric mass fraction
\begin{equation}
x(t) \equiv \frac{M_{\rm atm}(t)}{M_p(t)},
\end{equation}
where the total planet mass is
\begin{equation}
M_p(t) = M_{\rm core} + M_{\rm atm}(t),
\end{equation}
and the core mass $M_{\rm core}$ is fixed. In this case, atmospheric escape is driven by extreme ultraviolet stellar irradiation (XUV). Because mass is removed from both the atmosphere and the total planet mass, it follows that the evolution of $x$ is
\begin{equation}
\frac{\mathrm{d}x}{\mathrm{d}t}
=
-\frac{\dot M_{\rm esc}}{M_p}\,(1-x).
\end{equation}

The instantaneous energy-limited atmospheric mass-loss rate is 
\begin{equation}
\dot M_{\rm esc}
=
\eta\,
\frac{\pi R_{\rm XUV}^3\,F_{\rm XUV}(t)}
{G\,M_p\,K},
\end{equation}
where $\eta$ is the heating efficiency, $F_{\rm XUV}$ is the stellar XUV flux at the planet, and $K\le 1$ accounts for Roche-lobe effects of the reduction in binding energy where the atmosphere extends to a fraction of the planet's Hill radius \citep{Salz2016}.  The stellar XUV luminosity is parameterized by an initial saturated phase followed by power-law decay \citep[e.g.,][]{Rogers_Owen_2021},
\begin{align}
L_{\rm XUV}(t) &=
\begin{cases}
\varepsilon_{\rm XUV} L_\star, & t \le t_{\rm sat},\\[4pt]
\varepsilon_{\rm XUV} L_\star \left(\dfrac{t}{t_{\rm sat}}\right)^{-\alpha},
& t>t_{\rm sat},
\end{cases}
\end{align}
with the associated flux being
\begin{align}
F_{\rm XUV}(t) &=
\dfrac{L_{\rm XUV}(t)}{4\pi a^2}.
\end{align}

The Roche geometric factor is taken to be \citep{Salz2016}
\begin{align}
K &= 1 - \frac{3}{2\xi} + \frac{1}{2\xi^3},\\
\xi &\equiv \frac{R_L}{R_{\rm XUV}}
\end{align}
with
\begin{align}
R_L &\simeq a\left(\frac{M_p}{3M_\star}\right)^{1/3},
\end{align}
where $M_\star$ is the stellar mass and $a$ is the planet's semimajor axis. 

The XUV absorption radius is parameterized as
\begin{equation}
R_{\rm XUV} = R_p + n_H H
\end{equation}
where the scale height is given by
\begin{equation}
H =\frac{k_B T_{\rm therm}}{\mu}\frac{R_p^2}{G M_p}.
\end{equation}
Here $R_p$ is the photospheric radius, $T_{\rm therm}$ is a characteristic upper-atmosphere temperature, and $\mu$ is the mean molecular mass. The parameter $n_H$ is treated as an order-unity factor encapsulating thermospheric structure and opacity rather than as a literal number of physical scale heights. It is a simplification \citep[cf.][]{Salz2016} that is justified in so far as our results are not critically dependent on this value within plausible limits. 

The characteristic upper-atmosphere temperature $T_{\rm therm}$ is obtained from stellar properties by scaling it to the equilibrium temperature implied by those properties,

\begin{equation}
T_{\rm eq}
=
\left[
\frac{(1-A)\,L_\star}
{16\pi\sigma_{\rm SB}\,a^2\,f}
\right]^{1/4},
\end{equation}

\noindent where $A$ is the Bond albedo, $\sigma_{\rm SB}$ is the Stefan--Boltzmann constant, and $f$ accounts for heat redistribution (with $f=1$ for full-planet reradiation). The upper-atmosphere temperature is then taken to be

\begin{equation}
T_{\rm therm} = \alpha_T\,T_{\rm eq},
\end{equation}

\noindent where $\alpha_T \gtrsim 1$ is a dimensionless scaling factor that captures the prospects for additional thermospheric heating by XUV irradiation. In practice, $\alpha_T$ is treated as a fixed parameter. Varying $\alpha_T$ primarily rescales the default atmospheric scale heights and hence the effective XUV absorption radii, and therefore plays a role analogous to other escape-efficiency parameters such as $\eta$ and $n_H$.

Combining these elements, the evolution equation solved numerically is
\begin{align}
\frac{\mathrm{d}x}{\mathrm{d}t}
&=
-(1-x)\,
\frac{\eta\,\pi\,R_{\rm XUV}^3}{G\,M_p\,K}\,
\frac{L_{\rm XUV}(t)}{4\pi a^2},
\end{align}
with
\begin{align}
M_p &= \frac{M_{\rm core}}{1-x}.
\end{align}

\begin{figure*}[t]
\centering
   \includegraphics[width=0.80\textwidth]{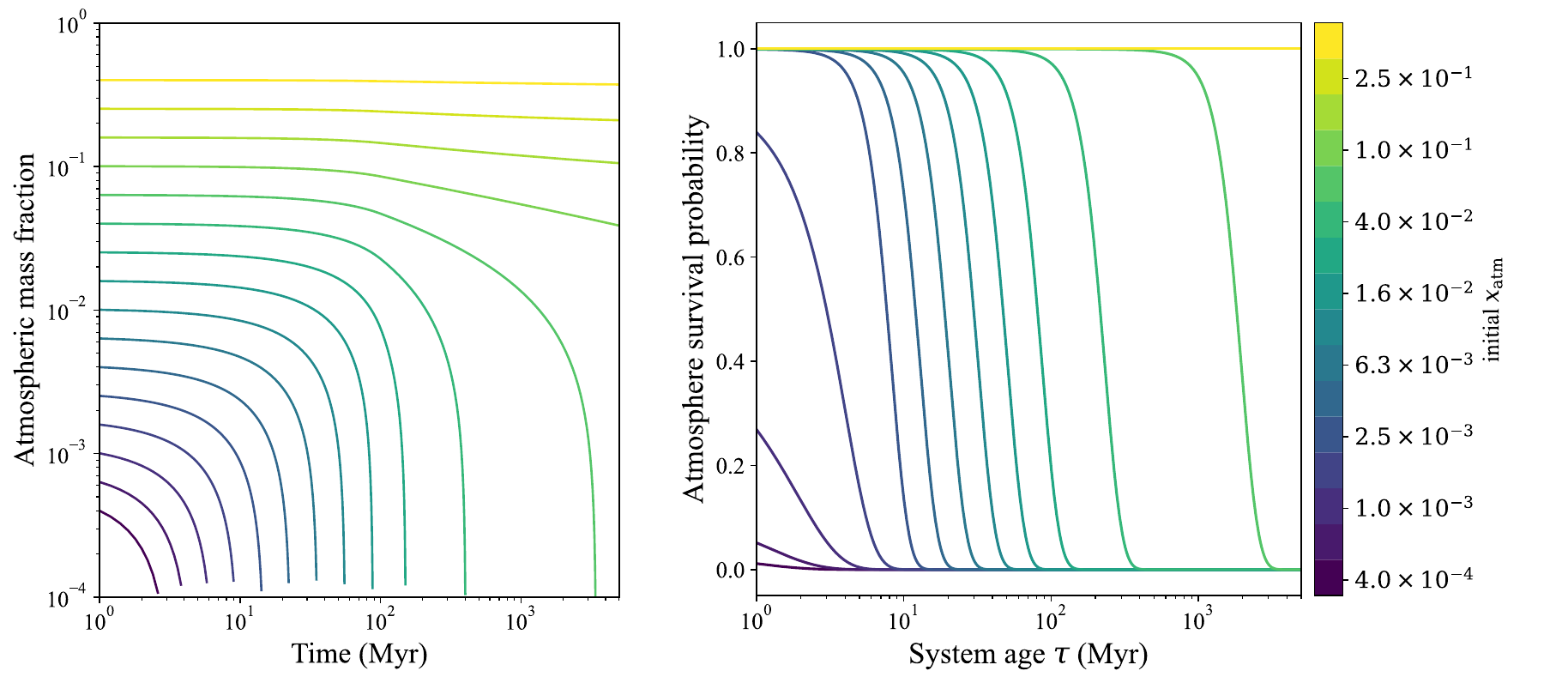}
    \caption{Atmosphere mass fractions as a function of time for planets each with a total mass of $5 M_\oplus$ and a period of 10 days. The panel on the left shows contours for different initial atmosphere mass fractions.  The panel on the right shows the associated atmosphere survival probabilities. Color bar for initial mass fractions of atmosphere applies to both plots. }
\label{fig:escape}
\end{figure*}

Example calculations based on the parameters in Table \ref{tbl:Table1}, in part from \cite{Salz2016} and \cite{Rogers_Owen_2021}, illustrate the sharp nature of energy-limited escape (Figure \ref{fig:escape}).  Planets either retain their atmospheres or they do not, with the time interval for partial escape being relatively limited for the timescales of order a Gyr of interest here.  Therefore, escape can be treated as a probability for atmospheric ``survival".

\subsubsection{Survival Probability}
\label{section:survival}

The atmospheric evolution is integrated forward in time using explicit timestepping:
\begin{align}
M_{\rm atm}(t+\Delta t) &=
M_{\rm atm}(t) - \dot M_{\rm esc}(t)\,\Delta t,
\end{align}
and therefore
\begin{align}
x(t+\Delta t) &=
\frac{M_{\rm atm}(t+\Delta t)}
{M_{\rm core}+M_{\rm atm}(t+\Delta t)}.
\end{align}
The characteristic stripping time $t_{\rm strip}$ is defined as the first time $x(t)$ falls below a threshold $x_{\rm thresh}$ that is near zero.

The default probability mapping converts $t_{\rm strip}$ into a smooth survival probability,
\begin{equation}
p_{\rm survive}(\tau)
=
\frac{1}{2}\,
\mathrm{erfc}\!\left(
\frac{\tau - t_{\rm strip} + \Delta}
{\sqrt{2}\,\sigma}
\right),
\end{equation}
with a temporal offset 
\begin{equation}
\Delta
=
\sqrt{2}\,\sigma\,
\mathrm{erfc}^{-1}(2\varepsilon),
\end{equation}
such that $p_{\rm survive}(\tau=t_{\rm strip})=\varepsilon$.

\begin{table}[h]
\centering
\caption{Atmospheric escape probability parameters used in models.}
\begin{tabular}{@{}lll@{}}
\toprule
Parameter & Definition & Adopted value \\
\midrule
$\eta$                    & Heating efficiency                          & $0.16$ \\
$\varepsilon_{\rm XUV}$   & Saturated XUV fraction                      & $10^{-3}$ \\
$t_{\rm sat}$             & XUV saturation time                         & $100\ \rm Myr$ \\
$\alpha$                  & Post-saturation decay index                 & $1.1$ \\
$n_H$                     & XUV absorption scale factor                 & $2.0$ \\
$\mu$                     & Mean molecular mass                         & $2.3\,m_H$ \\
$T_{\rm therm}$           & Upper-atmosphere temperature                & $3\,T_{\rm eq}$ \\
$x_{\rm thresh}$          & Stripping threshold                         & $10^{-4}$ \\
$\sigma$                  & Probability transition width                & $10\ \rm Myr$ \\
$\varepsilon$             & Survival probability at $t_{\rm strip}$     & $10^{-3}$ \\
$M_\star$                 & Stellar mass                                & $0.9M_{\odot}$ \\
$L_\star$                 & Stellar luminosity                          & $1.0L_{\odot}$ \\
\bottomrule
\end{tabular}
\label{tbl:Table1}
\end{table}

\subsubsection{Hydrodynamic Escape/Core-powered Mass Loss}
\label{section:hydrodynamic_escape}
The energy-limited photoevaporation atmosphere escape calculations can in some cases result in an
atmosphere that is unstable against hydrodynamic escape (Parker wind/Bondi escape). To account for this
we apply a threshold parameter that accounts for hydrodynamic escape in the form of the escape parameter $\lambda_{\rm rcb} = GM_p\mu / (k_B T_{\rm rcb} R_{\rm rcb})$ 
where $\mu$ is the mean molecular mass and
$T_{\rm rcb}$ is the temperature, each evaluated at the radial position of the radiative-convective boundary (RCB), $R_{\rm rcb}$. This serves to include core-powered mass loss implicitly \citep{ginzburg2016a, ginzburg2018a, gupta2019a} where the
planet's intrinsic cooling luminosity maintains an elevated $T_{\rm rcb}$ for Gyr
timescales, keeping $\lambda_{\rm rcb}$ small and the atmosphere vulnerable
to escape. In our models, $T_{\rm rcb}$ comes in part from the planet's intrinsic luminosity. The $\lambda_{\rm rcb}$ threshold therefore captures the essence of core-powered mass-loss without requiring an explicit integration of the coupled thermal
and mass-loss evolution. It operates as a safeguard beyond XUV escape alone that guards against implausible atmospheres.  

We evaluate the threshold value for $\lambda_{\rm rcb}$ based on the analysis of \cite{gupta2019a}. The mass-loss rate due to a Parker wind at the sonic point is

\begin{equation}
    \dot{M} = \pi G^2 M_p^2 \, P_{\rm rcb} \,
              \frac{\mu^{5/2}}{(k_B T_{\rm rcb})^{5/2}} \,
              e^{-\lambda_{\rm rcb}},
    \label{eq:mdot_bondi}
\end{equation}

\noindent
where $P_{\rm rcb}$ is the pressure at the radiative--convective
boundary, and, again,  $\lambda_{\rm rcb}$
is the escape parameter evaluated at the RCB.
Equation~(\ref{eq:mdot_bondi}) is obtained by substituting the sonic-point radius
$R_s = GM_p/(2c_s^2)$ and the ideal-gas relation
$\rho_{\rm rcb} = P_{\rm rcb}\mu/(k_BT_{\rm rcb})$ into the equation for Bondi escape.
The corresponding atmospheric loss timescale is
$t_{\rm loss} = x_{\rm atm} M_p / \dot{M}$, where  $x_{\rm atm} \equiv M_{\rm atm}/M_p$ is the mass fraction of the envelope (atmosphere).  Based on this definition for the timescale of atmosphere loss, we have

\begin{equation}
    t_{\rm loss} = \frac{x_{\rm atm} \, (k_B T_{\rm rcb})^{5/2}}
                        {\pi G^2 M_p \, P_{\rm rcb} \, \mu^{5/2}} \,
                   e^{+\lambda_{\rm rcb}}.
    \label{eq:tloss}
\end{equation}

\noindent
We require that the atmosphere survive at least one protoplanetary disc dispersal timescale,
$t_{\rm loss} \geq t_{\rm disc}$, and solve for $\lambda_{\rm rcb}$ based on that condition:

\begin{equation}
    \lambda_{\rm rcb} \;\geq\; \ln\!\left(
        \frac{t_{\rm disc} \, \pi G^2 M_p \, P_{\rm rcb} \, \mu^{5/2}}
             {x_{\rm atm} \, (k_B T_{\rm rcb})^{5/2}}
    \right).
    \label{eq:lambda_min}
\end{equation}

\noindent
Evaluating Equation~(\ref{eq:lambda_min}) at representative values
($t_{\rm disc} = 3$~Myr, $M_p = 1$--$20\,M_\oplus$, $x = 10^{-4}$--$10^{-1}$,
$P_{\rm rcb} = 10$~bar, $T_{\rm rcb} = 1020$~K, $\mu = 2.3\,m_H$) yields
$\lambda_{\rm min} \approx 22$--$32$, with an intermediate value of $\approx 26$ at
$M_p = 5\,M_\oplus$ and $x_{\rm atm} = 0.01$.
Because $\lambda_{\rm rcb}$ depends on the logarithm of the various terms, the threshold is
relatively insensitive to the precise values of the planet parameters: varying $t_{\rm disc}$
over 1--10~Myr, $M_p$ over 1--20~$M_\oplus$, $x$ over four decades, and
$T_{\rm rcb}$ over 300--2000~K shifts $\lambda_{\rm min}$ by at most a few
units in each case.
We therefore adopt $\lambda_{\rm rcb} = 20$ as a
conservative threshold, noting that no combination of physically plausible
parameters produces $\lambda_{\rm min}$ below about $ 22$.
Planets with escape parameters $< 20$ at the RCB have their atmospheres stripped in the model population.

\section{Monte Carlo Sampling}
\label{section:monte_carlo_sampling}

A synthetic population of planets based on the modeling described above
(contained in the Python code \texttt{Planet\_LAB},
\url{https://github.com/eyoungucla/Planet_Lab})
was obtained using a Monte Carlo sampling procedure.
Here the stochastic selection of input parameters and the probabilistic
assignment of observable states is described.

\subsection{Primary Parameter Sampling}

A total of 1012 planetary realizations were generated for the results reported here. Results are not sensitive to the precise number beyond $\sim 200$ draws.  Each realization is prescribed by the total planet mass ($M_p$), total weight percent H$_2$, the pressure at the binodal surface, and the equilibrium temperature. While we assign distinct equilibrium temperatures for the atmospheric loss calculations (\S \ref{sec:escape}), we chose to use a single equilibrium temperature,  $T_{\rm eq}=500$~K, for constructing the pre-escape planet models.  This value is taken to be representative for the range of periods of about 10 to 100 days.  In reality, $T_{\rm eq}$ can vary from about $1000$~K to $400$ K for a solar-mass star over this range in period, depending upon albedo. Applying individual $T_{\rm eq}$ values over this range in period has only negligible effects on our results. 

Planet masses are drawn from a log-uniform distribution:
\begin{equation}
1.7 \le M_p/M_\oplus \le 12,
\end{equation}
in order to obtain equal probability per decade in mass. 

The bulk hydrogen mass fraction prior to atmospheric loss is sampled from a truncated log-normal distribution.  This distribution type is motivated by expectations from agglomeration of protoplanets with hydrogen envelopes coupled with fractional losses of hydrogen during multiple accretion events \citep{Gibrat1931}; a log-normal distribution is an emergent property of stochastic assembly (Appendix B). The adopted distribution has a median value of $1.8$~wt\% and a logarithmic dispersion of $0.4$ dex, with fixed bounds of $0.05$ and $10$~wt\% to exclude unrealistically gas-poor or gas-rich cases. The physical implications of the distribution parameters is discussed in Appendix B.

The binodal surface pressure $P_s$ is drawn from a uniform distribution between $0.2$ and $5$~GPa for planets with $M_p < 3~M_\oplus$ and between $0.4$ and $5$~GPa for more massive planets. This weak mass dependence is the result of numerical considerations, and the results are not sensitive to the precise lower bounds, as described above.  In practice, for almost all realizations, the upper bound of $5$ GPa, where binodal temperatures are low, is never attained, as  \texttt{Planet\_LAB} reduces pressure if the total amount of H$_2$ is insufficient to avoid negative concentrations of hydrogen in the cores.  The practical upper limit is $4$ GPa (e.g., Figure \ref{fig:Evolution}).

\subsection{Metal-core Classification}

Each model planet realization is assigned a binary metal-core flag prior to forward modeling in order to decide whether the forthcoming model should include a separate, Fe-rich metal core or not (i.e., whether the planet is differentiated in the terrestrial sense).  The classification predictor was obtained by analysis of 218 random planet realizations that correlated the input variables with the phase equilibria described in \S \ref{section:miscibility} and \S \ref{section:Material_properties}.  Each of these random draws produced planets with a weight fraction of H$_2$ in the core, and one can show by comparisons with the ternary phase equilibria, that for core hydrogen concentrations $< 0.5\%$, a discrete metal phase is expected to form with a bulk density greater than the surrounding silicate, indicating a strong likelihood for metal-silicate differentiation. A logistic function was used as a classification predictor for whether a given set of model parameters leads to metal–silicate differentiation and the formation of a metal core,

\begin{equation}
p({\rm metal\,core})= \frac{1}{1+e^{-z}},
\label{eqn:core_preidctor}
\end{equation}

\noindent where $z$ is a linear predictor formed from the input parameters,

\begin{align}
z ={}& 5.4069
      - 0.5040\,(M_p/M_\oplus) \nonumber\\
    & - 3.7032\,\mathrm{(H_2/wt\%)}
      + 1.0793\,(P_s/{\rm GPa}).
\label{eqn:z}
\end{align}

\noindent The coefficients in Equation \ref{eqn:z} were determined by logistic regression applied to 218 exploratory model realizations using the Python statsmodels.api routine \texttt{Logit}, where the outcome variable was binary (metal core present or absent). Here,

\begin{equation}
z = \ln\!\left(\frac{p}{1-p}\right),
\end{equation}

\noindent so that $z=0$ corresponds to $p=0.5$. The regression therefore determines the hyperplane in parameter space at which the probability of forming a metal core is $50\%$. The resulting predictor reproduces the classifications of the training models with an accuracy better than $92\%$.

Planets with $z>0$ are classified as metal-core objects and the models are run with discrete metal cores using an Earth-like metal fraction of $0.33$.  Based on prior random model realizations, an empirical mapping between total hydrogen and the density deficit of the metal phase comprising the differentiated metal core is invoked for each model where differentiation is indicated. The linear mapping is  $\Delta\rho_{\rm metal}=0.315\,(\mathrm{planet\ \,H2\;wt\%})$, motivated by the ternary phase-equilibria trends explored in our preliminary calculations. We have, therefore, that $\rho_{\rm metal} = \rho_{\rm liquid\ Fe}(1-0.315(\mathrm{planet\ \,H2\;wt\%}))$ for our differentiated models.

\subsection{Assignment of Evolutionary Properties}
After producing a model planet, a second stochastic step assigns evolutionary and orbital properties intended to represent an observational population. For each planet realization, a stellar age $\tau$, orbital period $P_{\rm orb}$, and an initial atmospheric mass fraction parameter $x_0$ are sampled independently and assigned to the planet.

System ages are drawn from a kernel-density estimate constructed from stellar ages in the comparison sample extracted from the NASA Exoplanet Archive. Sampling is restricted to the interval
\begin{equation}
50 \le \tau \le 10^{4}~{\rm Myr},
\end{equation}
consistent with the empirical age distribution of observed systems.  Orbital periods are sampled from a log-uniform distribution between 7 and 100 days, consistent with the period range considered in the observational comparison.

The initial atmospheric mass fraction $x_0$, defined as the ratio of atmospheric mass to total planet mass at formation, is treated as a stochastic initial condition. Because the primordial envelope mass could plausibly correlate with the total hydrogen inventory of a planet, we also experimented with introducing a correlation between $x_0$ and the bulk H$_2$ mass fraction prior to atmospheric loss. Tests with correlation coefficients as large as $\rho = 0.9$ produced results that were statistically indistinguishable from those obtained with independent draws. We therefore adopt the simpler assumption of independent sampling in the models reported here.

Values of $x_0$ are drawn from a log-normal distribution with a median of $2\times10^{-2}$ (a trial value consistent with the median total hydrogen abundance) and a multiplicative one-sigma scatter of 2, anticipating order-of-magnitude variability in primordial envelope masses. The distribution is truncated to the interval $10^{-5}$--$10^{-1}$ to exclude unphysical values in this context. For all planet realizations the initial atmospheric mass fraction must satisfy $x_0 \ge x_{\rm pd}$, where $x_{\rm pd}$ is the modeled present-day atmospheric mass fraction.

\subsection{Atmospheric Survival as a Bernoulli Process}
Atmospheric retention is treated probabilistically. For each realization the atmosphere survival probability $p_{\rm survive}(\tau)$ is computed from the sampled parameters and planetary properties using the atmospheric escape framework described in \S~\ref{section:survival}. A Bernoulli trial is then performed such that planets retain their atmospheres with probability $p_{\rm surv}$ and are otherwise represented by their stripped cores.

An additional stability criterion is imposed such that models with $\lambda_{\rm rcb}<20$ are always classified as stripped, corresponding to instances where envelopes are dynamically unstable (see \S \ref{section:hydrodynamic_escape}).

\subsection{Resulting Population}
The Monte Carlo realizations comprise an ensemble of model planets conditioned on physically motivated priors and filtered by probabilistic atmospheric survival, enabling direct comparison with observed mass--radius and period distributions.

\section{Results}
\label{section:results}

\subsection{Model Planets}
\begin{figure*}[t]
\centering
\begin{minipage}[t]{0.47\textwidth}
    \centering
    \includegraphics[width=\textwidth]{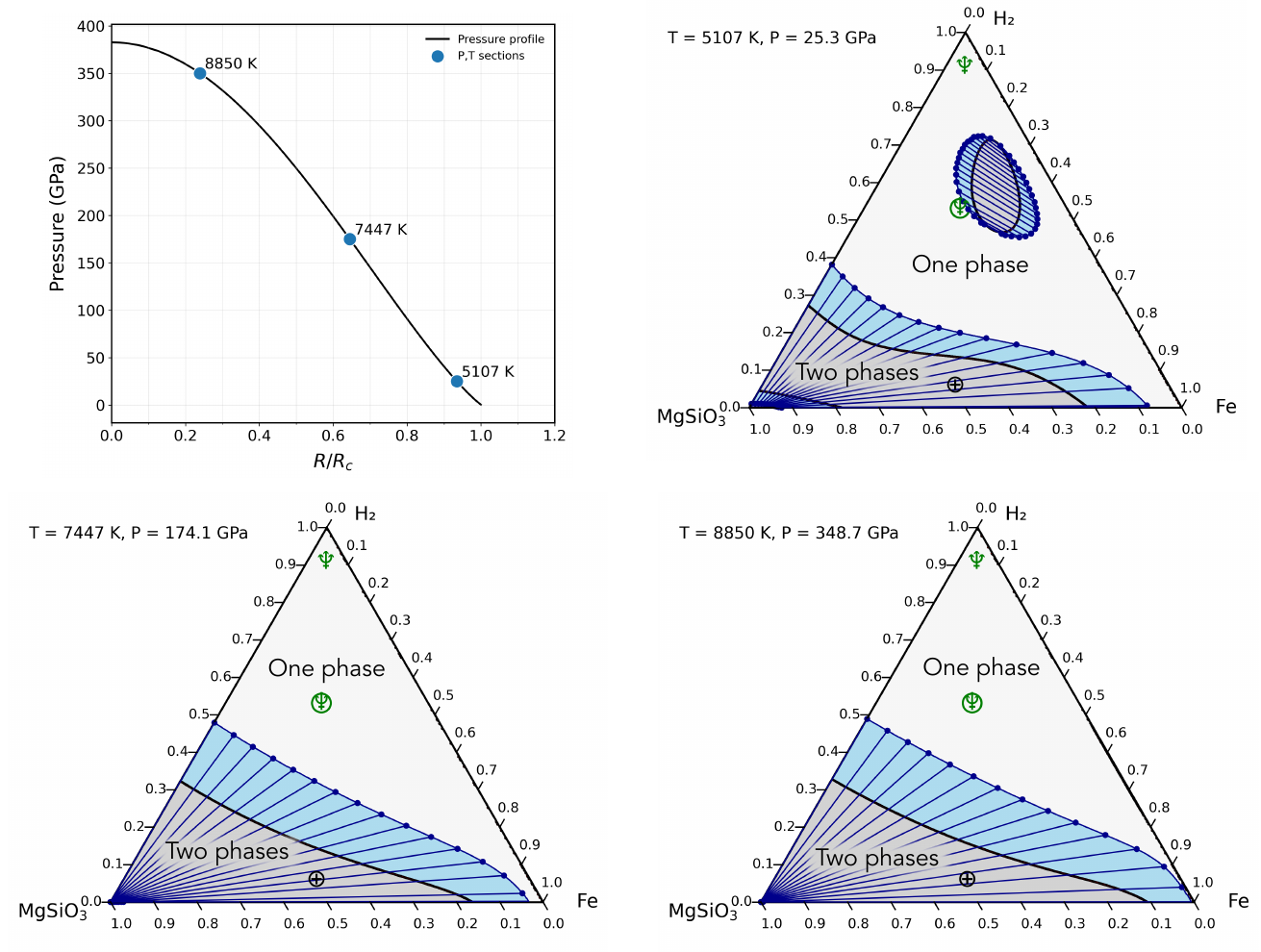}
    \caption{Phase diagrams for the interior of a model 6$M_\oplus$ planet 
    that has $3$ wt\% total hydrogen with a binodal pressure of $0.3$ GPa 
    and $L_{\rm int} = 3.3\times 10^{16}$ W. The upper left panel shows 
    pressure vs depth in the molten interior, and positions for the ternary 
    phase diagrams for three different depths in the planet. Ternary 
    coordinates are in mole fractions. Reference bulk compositions projected 
    into this ternary space are shown for Earth ($\oplus$), the model 
    sub-Neptune interior 
    (\includegraphics[height=1em]{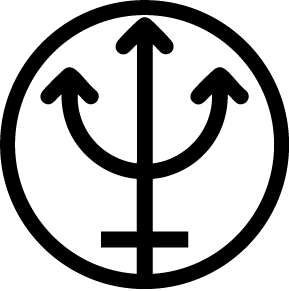}), and Neptune 
    (\includegraphics[height=1em]{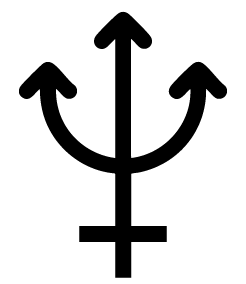}).}
    \label{fig:ternary_pt3GPa}
\end{minipage}
\hfill
\begin{minipage}[t]{0.47\textwidth}
    \centering
    \includegraphics[width=\textwidth]{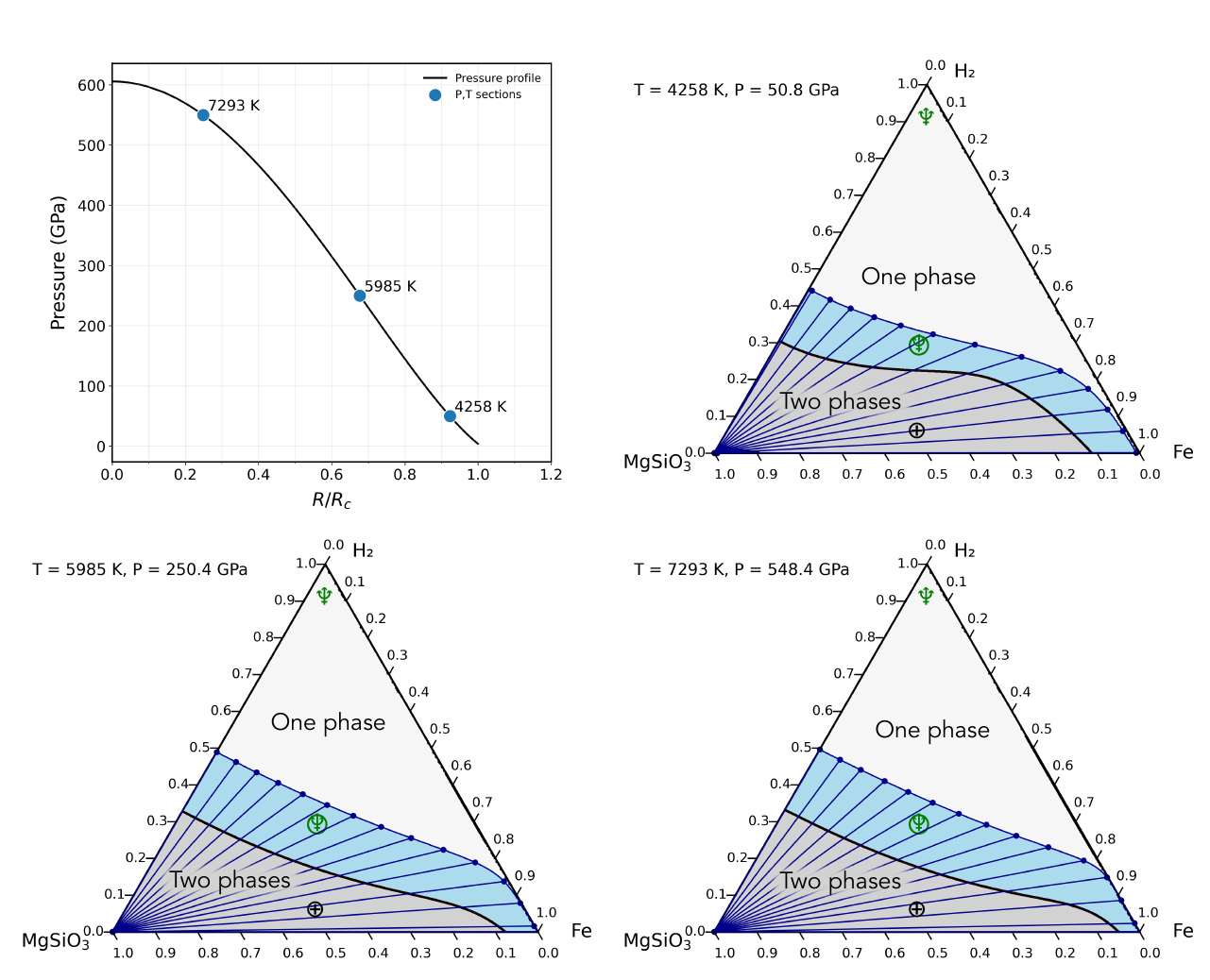}
    \caption{Phase diagrams for the interior of a model 6$M_\oplus$ planet 
    that has $3$ wt\% total hydrogen with a binodal pressure of $4$ GPa 
    and $L_{\rm int} = 1.6\times 10^{15}$ W. The upper left panel shows 
    pressure vs depth in the molten interior, and positions for the ternary 
    phase diagrams for three different depths in the planet. Ternary 
    coordinates are in mole fractions. Reference bulk compositions projected 
    into this ternary space are shown for Earth ($\oplus$), the model 
    sub-Neptune interior 
    (\includegraphics[height=1em]{sub-neptune.png}), and Neptune 
    (\includegraphics[height=1em]{neptune.png}).}
    \label{fig:ternary_4GPa}
\end{minipage}
\end{figure*}

Two examples of the interiors of model planets are shown in Figures \ref{fig:ternary_pt3GPa} and \ref{fig:ternary_4GPa}, representing early and late stages of evolution. The figures show pressure vs depth in the molten interior, and three phase diagrams at three representative depths.  In both cases, the planets are 6$M_\oplus$ with $3$ wt\% total hydrogen. Figure \ref{fig:ternary_pt3GPa} shows the state early, with a binodal pressure, $P_s$, of $0.3$ GPa, while Figure \ref{fig:ternary_4GPa} shows a later stage of evolution where $L_{\rm int}$ has decreased from the initial value of $3.3\times 10^{16}$ W in Figure \ref{fig:ternary_pt3GPa} to $1.6\times 10^{15}$ W and the binodal pressure has increased accordingly to 4 GPa.  

At this total hydrogen mass fraction, the core of this planet is predicted to be essentially one phase throughout its history. Inspection of the core compositions in ternary space in the two figures shows that exsolution of H$_2$ from the core with cooling lowers the core H$_2$ fraction, placing it on the edge of the two-phase field.  However, application of the lever rule indicates that the core is almost entirely composed of the phase at the high-hydrogen end of the tie line passing through the core composition.  This phase is nothing like an Fe core in the terrestrial sense, but instead is composed of 0.3 MgSiO$_3$, 0.35 Fe, and 0.35 H$_2$ by mole, or 60\% MgSiO$_3$, $38.5\%$ Fe, and $1.5\%$ H$_2$ by mass (based on simple conversion using molecular weights). This is an example of an undifferentiated planet in our model architecture. 

Also shown in Figures \ref{fig:ternary_pt3GPa} and \ref{fig:ternary_4GPa}, is bulk Earth with $0.17$ weight \% H$_2$ \citep[e.g.,][]{Young_Nature_2023} showing coexistence of a silicate (Earth's mantle) and an H-bearing Fe metal phase (Earth's metal core).  Planets between the 3\% for these sub-Neptune models and Earth will have core hydrogen concentrations intermediate between these endmembers, with $0.5\%$ by mass being the cutoff for silicate-metal differentiation, as prescribed by the slopes of the tie lines between the silicate and metal phases and implemented using Equation \ref{eqn:z}.

\subsection{Reference Planets in Mass--Radius Space}
\label{section:reference_planets}

\begin{figure*}
\centering
\begin{minipage}[t]{0.48\textwidth}
    \centering
    \includegraphics[scale=0.47]{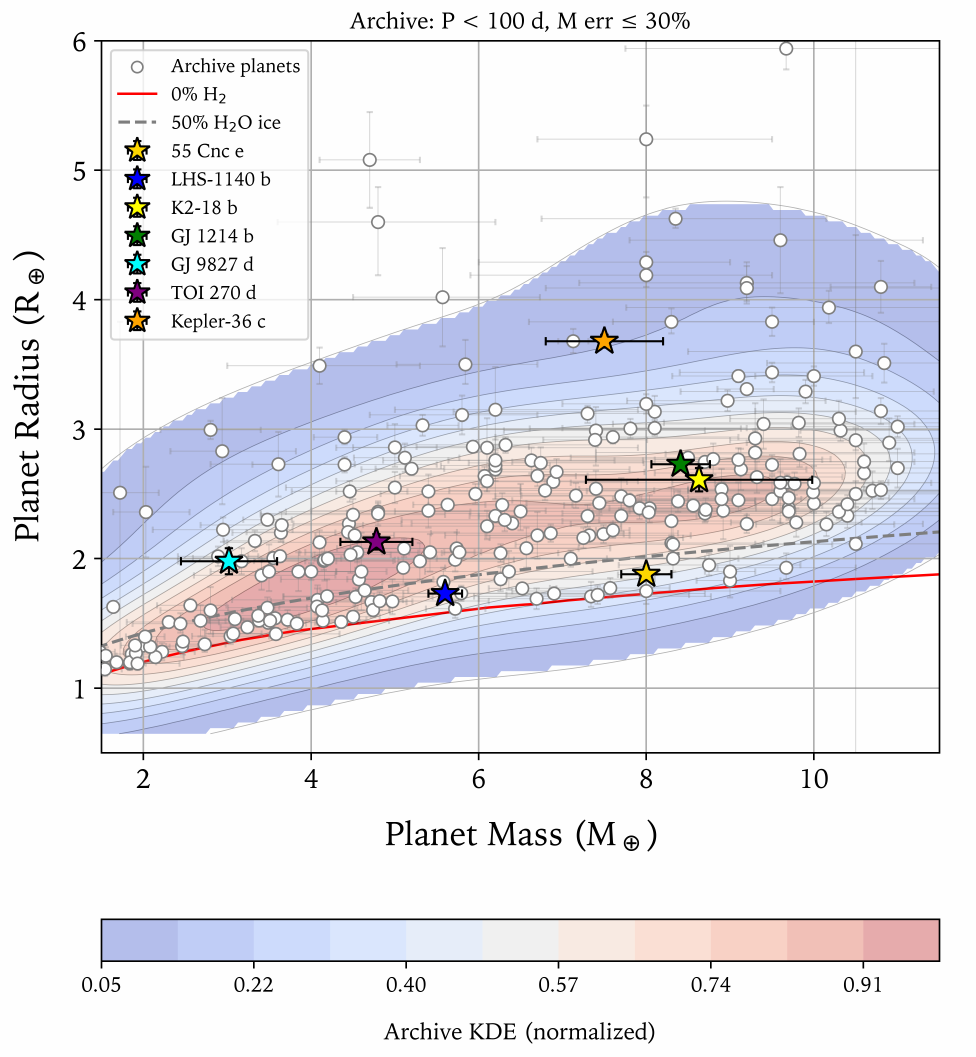}
    \caption{Mass versus radius diagram for 261 archive planets, together with several example planets discussed in the text. Planets all have periods $< 100$ days and mass precision of less than or equal to 30\%. Contours are for equal kernel density, illustrating the density of points. No points are shown below the pure Earth-like mass vs.\ density curve since these planets are not included in our models. Also shown is a curve for 50\% by mass water ice, based on an Earth-like core mixed with low-pressure water ice, cf.\ \cite{Aguichine2021}.}
    \label{fig:Mass_vs_Radius_REFERENCE}
\end{minipage}
\hfill
\begin{minipage}[t]{0.48\textwidth}
    \centering
    \includegraphics[scale=0.47]{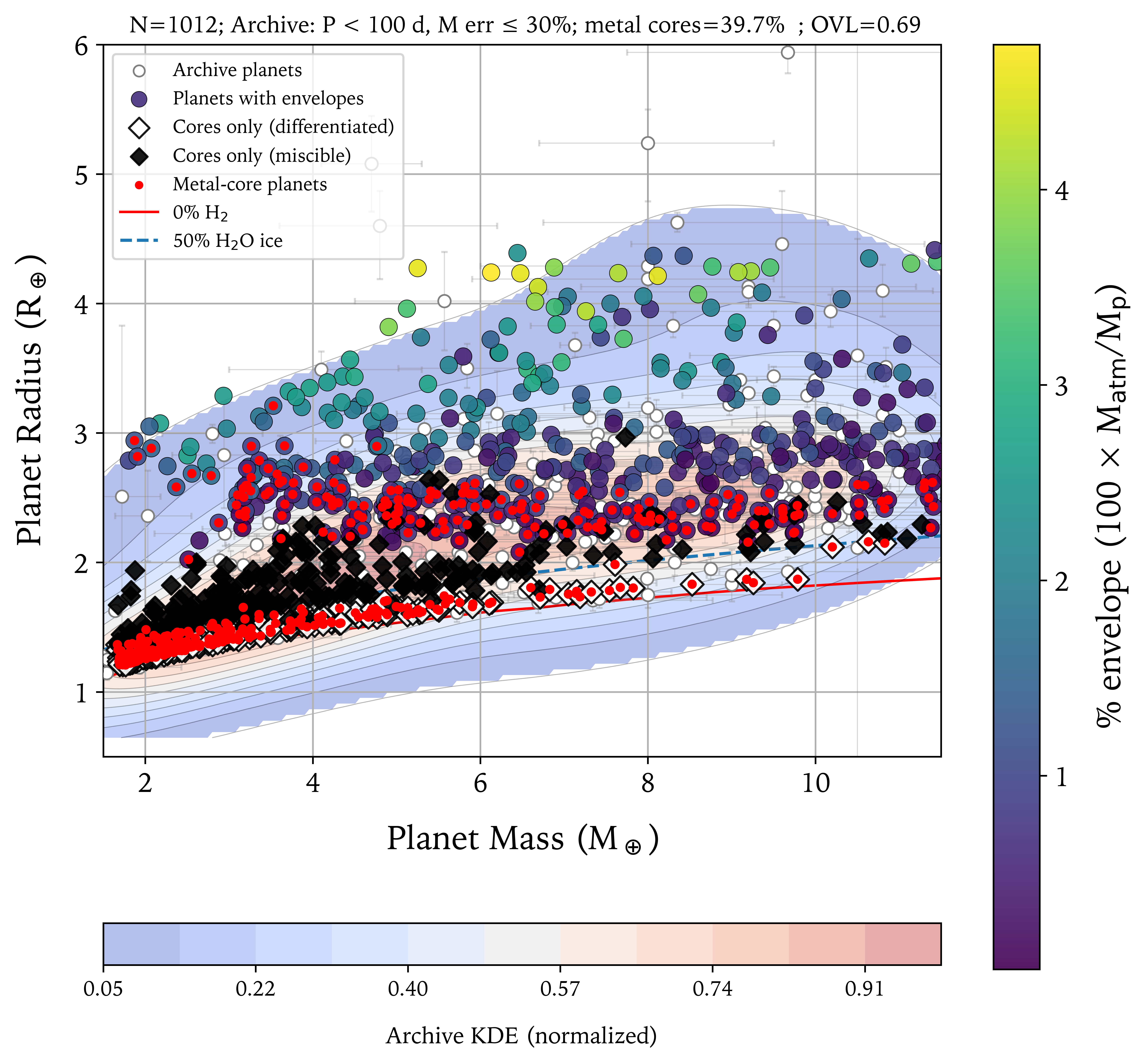}
    \caption{Comparison of model planets with observed planets in mass vs.\ radius space. Archive data, shown as white circles with error bars, define contours of equal kernel density (KDE contours). The colors of the model data points reflect their mass fractions of envelopes (in percent), with the exception of the white diamonds. Colored circles with no red dot have entirely miscible interiors with H$_2$-rich envelopes. Colored circles with red dots have H$_2$-rich envelopes and have discrete, Fe-rich metal cores. Black diamonds are fully miscible interiors that no longer have envelopes of hydrogen. White diamonds with red dots are differentiated interiors with Fe-rich metal cores that have been stripped of their primary H$_2$-rich envelopes.}
    \label{fig:Mass_vs_Radius}
\end{minipage}
\end{figure*}

The reference mass versus radius diagram for the 261 archive planets in this study that result from filtering for periods less than 100 days and mass precision of $\le 30\%$ is shown in Figure \ref{fig:Mass_vs_Radius_REFERENCE}. Contours are for equal kernel density of points. No points are shown below the pure Earth-like mass vs.\ radius curve since these planets are not included in our models.  A curve for 50\% by mass water ice, based  on an Earth-like core mixed with low-pressure water ice \citep[\textit{cf.},][]{Aguichine2021}, is shown for reference.  

\begin{figure*}
\centering
   \includegraphics[width=0.85\textwidth]{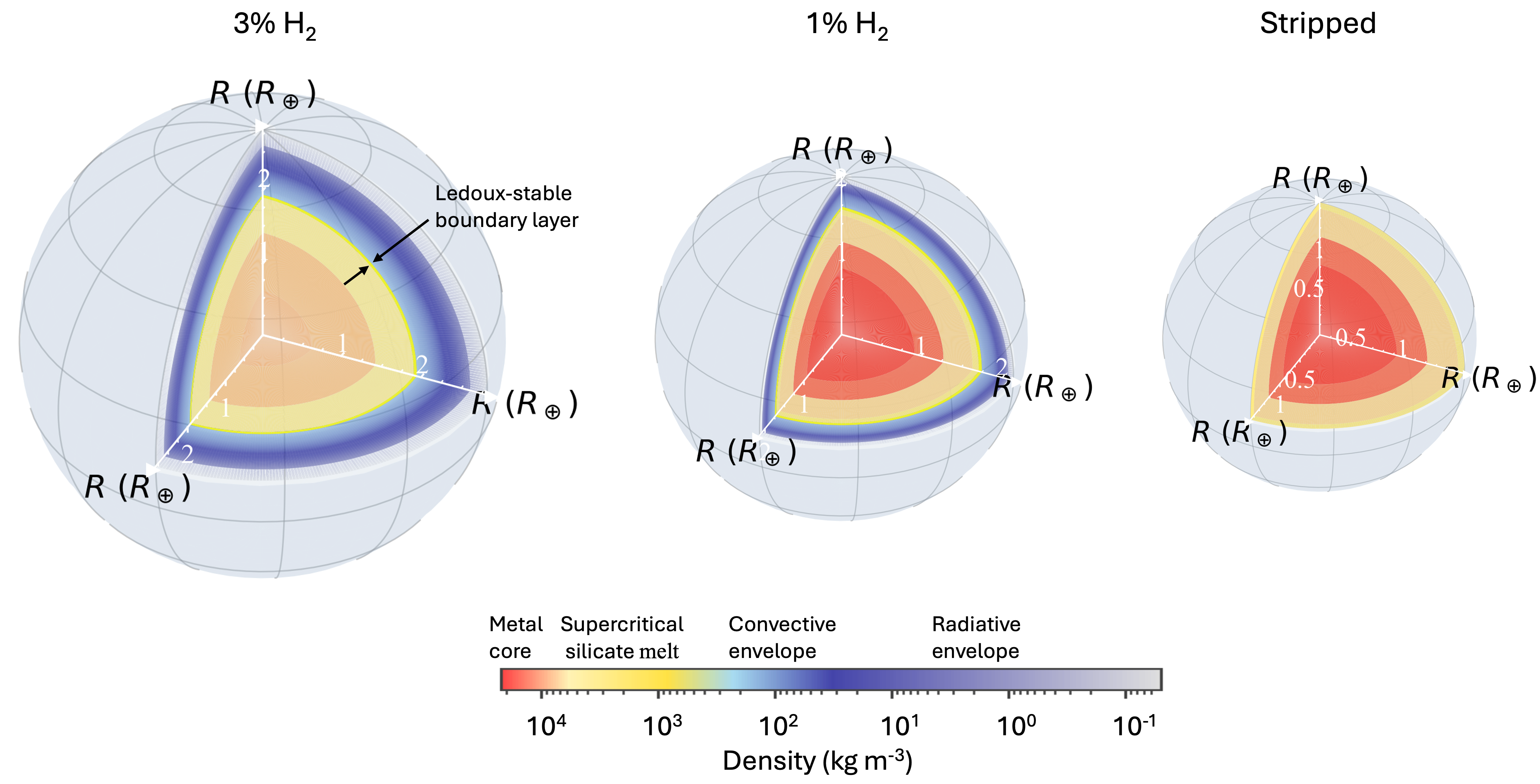}
    \caption{Three model planets illustrating a range of structures referred to in the text. Each planet has an initial mass of $6M_\oplus$ and a binodal pressure of $3.5$ GPa. The bulk fraction of H$_2$ by mass is indicated.  The planet on the right is the same as the $1\%$ H$_2$ case but with the atmosphere stripped. The Ledoux-stable boundary layers are shown to scale, represented by the yellow bands at the base of the H$_2$-rich envelopes.  Density contours indicate metal core, miscible silicate-H$_2$-Fe interior, and H$_2$-rich envelopes.} 
\label{fig:schematic}
\end{figure*}

Different type of planets in this population are exemplified by the seven specific examples shown in Figure \ref{fig:Mass_vs_Radius_REFERENCE}. We can use these planets to define roughly four types of super-Earths and sub-Neptunes for comparisons with our models. Planet 55 Cancri e is near the super-Earth/sub-Neptune boundary with an orbital period of about 1 day and an equilibrium temperature of 2400 K.  It is almost certainly  a lava world with a secondary atmosphere \citep{Hu2024_55Cnce}, and serves as a fiducial marker for such cases here. Similarly, planet LHS-1140 b lies in the transition between super-Earths and sub-Neptunes. It is tidally locked with an orbital period of about 25 days and an equilibrium temperature of 230 K.  It has been used as an example of a water world, with liquid water at the substellar point and perhaps a secondary atmosphere \citep{Cadieux2024_LHS1140b}. Planets TOI-270 d and GJ 9827 d have lower bulk densities, placing them in the regime of volatile-rich sub-Neptunes \citep{Benneke2024_TOI270d, Piaulet2024_GJ9827d}.  Water has been detected in the atmospheres of both planets, albeit at different equilibrium temperatures. GJ 9827 d may host a high–mean molecular weight, steam-rich atmosphere \citep{Piaulet2024_GJ9827d}, while JWST observations of TOI-270 d reveal a hydrogen-dominated atmosphere with prominent absorption from CH$_4$, CO$_2$, and likely H$_2$O and an inferred super-solar atmospheric metallicity \citep{Benneke2024_TOI270d}. Additional analyses report evidence for sulfur-bearing species in TOI-270 d’s atmosphere \citep{felix_competing_2025}. The atmospheric properties of TOI-270 d have been interpreted either as signatures of interaction with an underlying magma ocean \citep{Nixon2025_TOI270d} or as consistent with volatile-rich envelope scenarios, including liquid water configurations \citep{Holmberg2024_TOI270d}. Interior structure analyses further explore whether such water-rich states are thermodynamically consistent with the observed bulk density and atmospheric composition \citep{rigby_surface_2025}. Planets GJ 1214 b and K2-18 b are squarely in the sub-Neptune field in mass vs.\ radius space. GJ 1214 b is a ``warm" sub-Neptune with a period of just $1.6$ days and an equilibrium temperature of $\sim$500--600 K. It is thought to have a high-molecular-weight atmosphere, perhaps dominated by water, with a photochemical aerosol-rich cloud structure \citep{Kempton2023_GJ1214b, Gao2023_GJ1214b}. K2-18 b, with an orbital period of $33$ days around an M dwarf, exhibits a hydrogen-rich atmosphere with robust detections of CH$_4$ and CO$_2$, and stringent upper limits on NH$_3$ abundance \citep{Madhusudhan2023_K2-18b,hu_water-rich_2025}. Similar to TOI-270 d, these spectral constraints have been interpreted as consistent with either a relatively shallow H$_2$-rich envelope potentially overlying a volatile-rich interior \citep[the “Hycean” scenario;][]{Madhusudhan2021_Hycean}, or with a thicker H$_2$ envelope in contact with a magma ocean \citep{Shorttle_2024}.  Lastly, planet Kepler-36 c serves as a fiducial sub-Neptune with an H$_2$/He-rich envelope, evidenced by its low density of just 0.87 g cm$^{-3}$ \citep{Carter2012_Kepler36}.  It orbits its star with a period of 16 days.  

\subsection{Random Draws}
\label{setion:random_draws}
The 1012 random planets obtained in this study are plotted in mass vs.\ radius space in Figure \ref{fig:Mass_vs_Radius} together with the archive planets over the same mass range, shown also in Figure \ref{fig:Mass_vs_Radius_REFERENCE}.  The overall distributions are similar by eye, with all but three of the model planets falling inside of the 5\% normalized KDE contour for the archive data. The model planets concentrate in the same regions where the actual data are most prevalent (this is quantified below). Overall, 40\% of the model planets have metal cores, and the vast majority of these have radii $< \sim 2.5 M_\oplus$. 

The models produce four classes of planets.  The least dense planets have relatively large mass fractions of hydrogen-rich envelopes, and thus larger radii (circles with colors representing mass fractions of envelope in Figure \ref{fig:Mass_vs_Radius}).  These planets have fully miscible interiors, and thus no discrete, terrestrial-like metal cores.  Related to these are hydrogen-rich planets that have hydrogen-rich envelopes and also have Fe-rich metal cores (cores are denoted by red dots in Figure \ref{fig:Mass_vs_Radius}). As one expects, these planets have somewhat smaller radii for a given mass.  The third type of planet are those that are sufficiently hydrogen-rich that their interiors are fully miscible, but they have been stripped of their outer envelopes of hydrogen (black diamonds in Figure \ref{fig:Mass_vs_Radius}). The fourth planet type consists of those that have discrete Fe-rich iron cores and no longer have discernble hydrogen-rich envelopes (white diamonds in Figure \ref{fig:Mass_vs_Radius}). These are closest to Earth-like in structure and chemical makeup. Examples of model planet structures are shown in Figure \ref{fig:schematic}.

Broadly, the four groups cluster in ways analogous to the planet types described in \S \ref{section:reference_planets}. The model terrestrial-like super-Earths fall between the reference 0\% rock/metal and 50\% water curves in Figure \ref{fig:Mass_vs_Radius}. The region occupied by planets transitional between super-Earths and sub-Neptunes is populated by model planets with hydrogen-rich, fully miscible interiors stripped of their envelopes, and by planets with H$_2$-rich envelopes surrounding silicate and metal interiors with discrete Fe-rich metal cores. Model planets with fully miscible cores and significant H$_2$-rich envelopes occupy the same mass--radius space as canonical sub-Neptunes with H$_2$/He-rich atmospheres.

The model and observed planet radii are compared in Figure \ref{fig:R_vs_P_histogram}.  The latter have been corrected for completeness (See Appendix A) to bring the radius valley into relief, and to facilitate comparison between the unbiased random samples and the actual data. The model data exhibit a probability density distribution very similar to the archive data corrected for observational biases, with peaks and a pronounced radius gap at the same radii.  These features are mainly an inexorable result of atmosphere stripping.   

The radius gap is known to decrease in radius with increasing period, and this feature is often used as an arbiter for successfull models of planet formation and evolution \citep[e.g.,][]{owen2017a,gupta2019a}.  Model planets are compared with one estimate for the position of the radius gap as a function of orbital period by \cite{vaneylen2018a} in Figure \ref{fig:R_vs_period_with_gap}. The model data exhibit a gap in radius that generally decreases with orbital period. This is a qualitative result, but is generally consistent with expectations from the actual data. 

\subsection{Distribution Comparison Metrics}

In order to quantify the "chi-by-eye" apparent similarities between the synthetic planet population
and the observed archive sample in mass--radius space, we use
two statistics: a density overlap comparison
using kernel-density estimates (KDEs) and an 
energy-distance test calculated from the data points themselves.

The KDE overlap test provides a quantification of the visual overlap between the data sets \citep{Eidous2024}. The overlap coefficient, with range from 0 to 1, is defined as

\begin{equation}
\mathrm{OVL}
=
\int \min\!\left[f(\mathbf{x}),\,g(\mathbf{x})\right]\, d\mathbf{x},
\end{equation}
where $f(x)$ and $g(x)$ denote the KDEs of the model and archive
distributions, respectively, and min in this context refers to the minimum of the two values (the overlap can not exceed either distribution's density at $x$). The two populations here are the archive data (A) and the model data (M).

\begin{figure*}[t!]
\centering
\begin{minipage}[t]{0.47\textwidth}
    \centering
    \includegraphics[width=\textwidth]{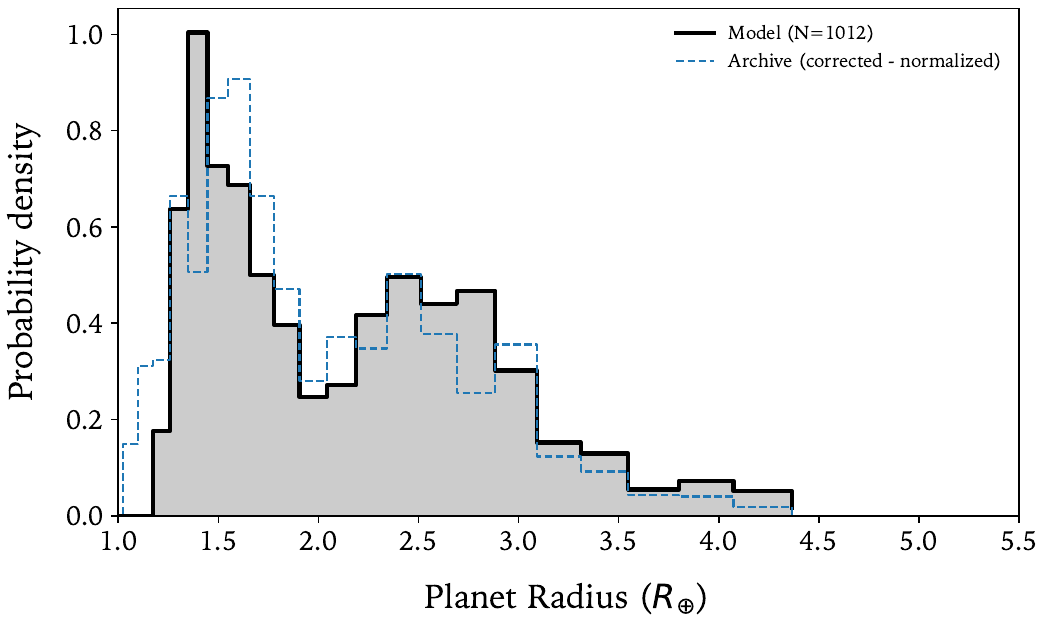}
    \caption{Histograms comparing model planet (grey filled) and observed 
    planet radii (dashed line). The latter are corrected for observational 
    biases (See Appendix A}). 
    \label{fig:R_vs_P_histogram}
\end{minipage}
\hfill
\begin{minipage}[t]{0.47\textwidth}
    \centering
    \includegraphics[width=\textwidth]{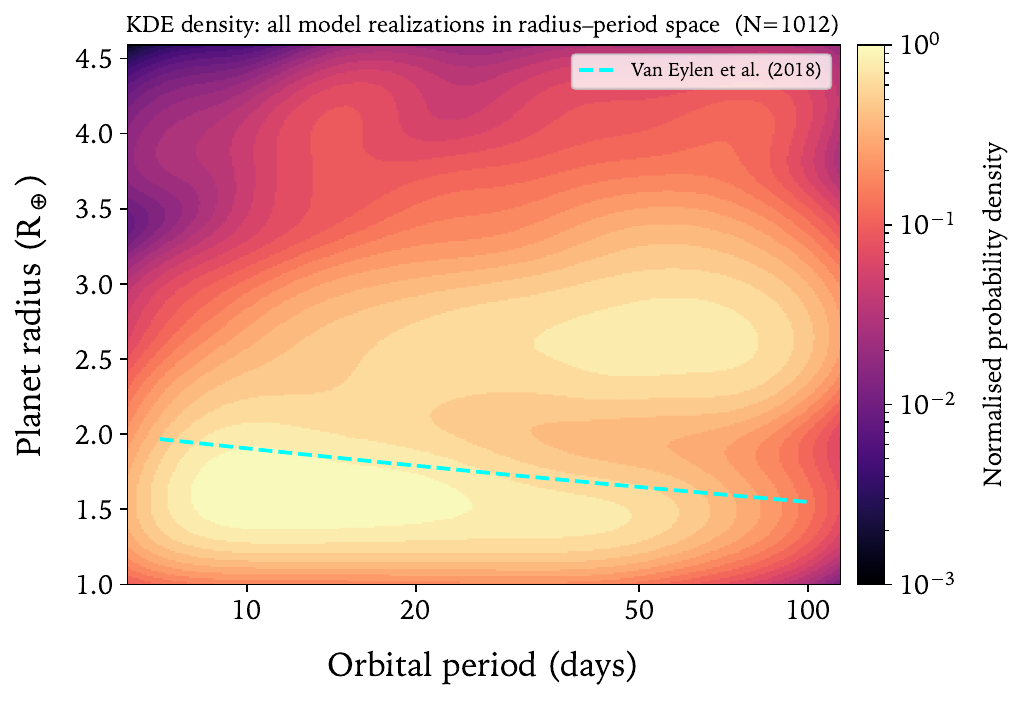}
    \caption{Kernel density estimate plot of model planets in radius vs.\ period space showing a 
    decrease in the model radius gap with orbital period. The position of 
    the actual radius gap ca.\ 2018 as determined by \cite{vaneylen2018a} is shown 
    for reference.}
    \label{fig:R_vs_period_with_gap}
\end{minipage}
\end{figure*}

The overlap coefficient measures the fraction of shared probability
mass between two distributions, ranging from zero for no overlap
to unity for identical distributions. It is sensitive to overall spread and to various peaks and valleys in the distributions.
Because finite sampling and observational scatter limit the maximum
attainable overlap, we estimate an empirical reference value
$\mathrm{OVL}(A,A)$ by computing the overlap between random halves
of the archive planet data themselves. 
We then define the normalized overlap as the ratio
\begin{equation}
R_{\rm OVL}=\frac{\mathrm{OVL}(\rm M,A)}
     {\mathrm{OVL(A,A)}},
\end{equation}
which measures the fraction of observationally achievable overlap
reproduced by the model population in view of the scatter among the data.

In this case, we obtain $\mathrm{OVL}(M,A)=0.69$
and $\mathrm{OVL(A,A)} = 0.82$,
implying $R_{\rm OVL}\approx0.84$.
The model planets therefore reproduce 84\% of the similarity achievable given the scatter in the actual planet data. 

In order to further explore the statistical significance of the $\sim 20\%$ disparity between the archive data and the model data, we use their energy distance.  The energy distance between two distributions is a scalar measure of their dissimilarity, equal to zero if and only if the distributions are identical, and increasing as they diverge. The name draws on the analogy with electrostatic potential energy, for example: just as the energy required to separate two charged bodies depends on the distance between them, the energy distance between two distributions is constructed from the mean distances between points drawn from each in a ``statistical potential energy". Energy distance can be estimated directly from pairwise point distances without assuming any parametric form for the underlying distributions. We compute the energy distance between
the model and archive samples $\{M_i\}$ and $\{A_j\}$ as follows  \citep{Rizzo2016}:
\begin{equation}
\mathcal{E}(M,A)
=
2\,\mathbb{E}\,\|M-A\|
-
\mathbb{E}\,\|M-M'\|
-
\mathbb{E}\,\|A-A'\|,
\end{equation}
where $\mathbb{E}$ denotes the expected value and $\|M-A\|$ is the Euclidean distance over all pairs; the expectations are 
averages over all pairwise Euclidean distances between
sample points.
The statistic vanishes entirely only when the two samples are drawn from the
same underlying distribution.
Statistical significance is assessed using a permutation test in which the identities of data points are shuffled randomly for the pooled model-archive data.  Each shuffled set is split into two sets and a value for $\mathcal{E}$ is evaluated, yielding the set $\mathcal{S} = \{\mathcal{E}_1, \mathcal{E}_2, \ldots,\mathcal{E}_N\}$. The empirical p-value under the null hypothesis of a common underlying distribution for the model and archive data, representing the fraction of permuted statistics in $\mathcal{S}$ that exceed $\mathcal{E}_{\rm obs}$, is then \citep{Phipson2010}

\begin{equation}
p = \frac{1 + \sum_{k=1}^{N}\mathbf{1}[\mathcal{E}_k > \mathcal{E}_{\rm obs}]}{1 + N}
\end{equation}

\noindent where $\mathbf{1}[\mathcal{E}_k > \mathcal{E}_{\rm obs}]$ is 1 if the condition is satisfied and otherwise zero, and $\mathcal{E}_{\rm obs} \equiv \mathcal{E}(M,A)$.  

We obtain $\mathcal{E}(M,A)=0.05$ with
$p\simeq2\times10^{-3}$ when comparing the model planets and the actual data ($p$ is small but also resolution limited, and thus depends on $N$, a consequence of the small value for $\mathcal{E}(M,A)$).  These values indicate that the two populations are statistically distinguishable (the null is rejected given the small $p$) despite the similarity in their distributions visible in Figure \ref{fig:Mass_vs_Radius}.

Together these tests indicate a substantial overlap between the model planets and the archive data, including the various peaks and valleys in the distributions, but because the samples are large enough to detect differences to high confidence, the small differences between the distributions are statistically resolved.  We did not correct the archive data for observatonal biases in mass-radius space, because of the uncertainties involved, so some of the remaining differences are surely due in part to comparing data with no observational biases (models) to data with observational biases (archive). 

\section{Discussion}
\label{discussion}

\subsection{Planet Types}
The randomly selected set of planets with structures governed by the degree of miscibility between silicates, iron metal, and hydrogen span the range of planet types exhibited by sub-Neptunes and super-Earths. To the extent that the model and archive planet populations are similar in their distribution in mass-radius and radius-period space, one can infer that miscibility of hydrogen may be an important determining  factor in planet formation in general over the relevant mass range of about 2 to 10 $M_\oplus$.  

The models are necessarily incomplete. For example, we have reduced the complex chemistry in these planets to a ternary system.  In reality, the abundant hydrogen in their interiors will result in the formation of water or water-like species in the supercritical melts \citep{Ikoma2006,Kite2021,Schlichting_Young_2022} and this water is expected to exchange with the dense H$_2$-rich envelopes at equilibrium. The presence of water, while of great importance, is not expected to change the overall structure of our model planets.  The {\it ab initio} molecular dynamics calculations upon which the hydrogen miscibility is based indeed demonstrate the conversion of H$_2$ to H$_2$O and OH-like species \citep{gilmore_core-envelope_2025} in the supercritical melt phase, consistent with macroscopic thermodynamic predictions at these conditions \citep{Schlichting_Young_2022, werlen_sub-neptunes_2025}. Therefore, the existence of water in the interiors is ``baked in" to our calculations. Water that partitions from the supercritical melt to the hydrogen-rich envelope would slightly alter the mean molecular weight of the gas, but the change would be relatively minor (e.g., H$_2$O replacing $1/2$ O$_2$).

In the planets where the envelopes are stripped, we have not included their subsequent evolution, implicitly assuming that their interiors become isolated with their full complement of hydrogen. We have therefore tacitly assumed that once the atmospheres are removed, rapid cooling quenches the system, preserving the H$_2$ inherited during the magma ocean phase.  However, alternatively, cooling may be sufficiently slow that hydrogen degassing can occur before quenching \citep[e.g.,][]{Rogers_2025}. In addition, we do not try to model the geological evolution of planets stripped of their envelopes that leads to high-molecular-weight secondary atmospheres, including outgassing of water produced in their interiors. 

\begin{figure}
\centering
   \includegraphics[width=0.48\textwidth]{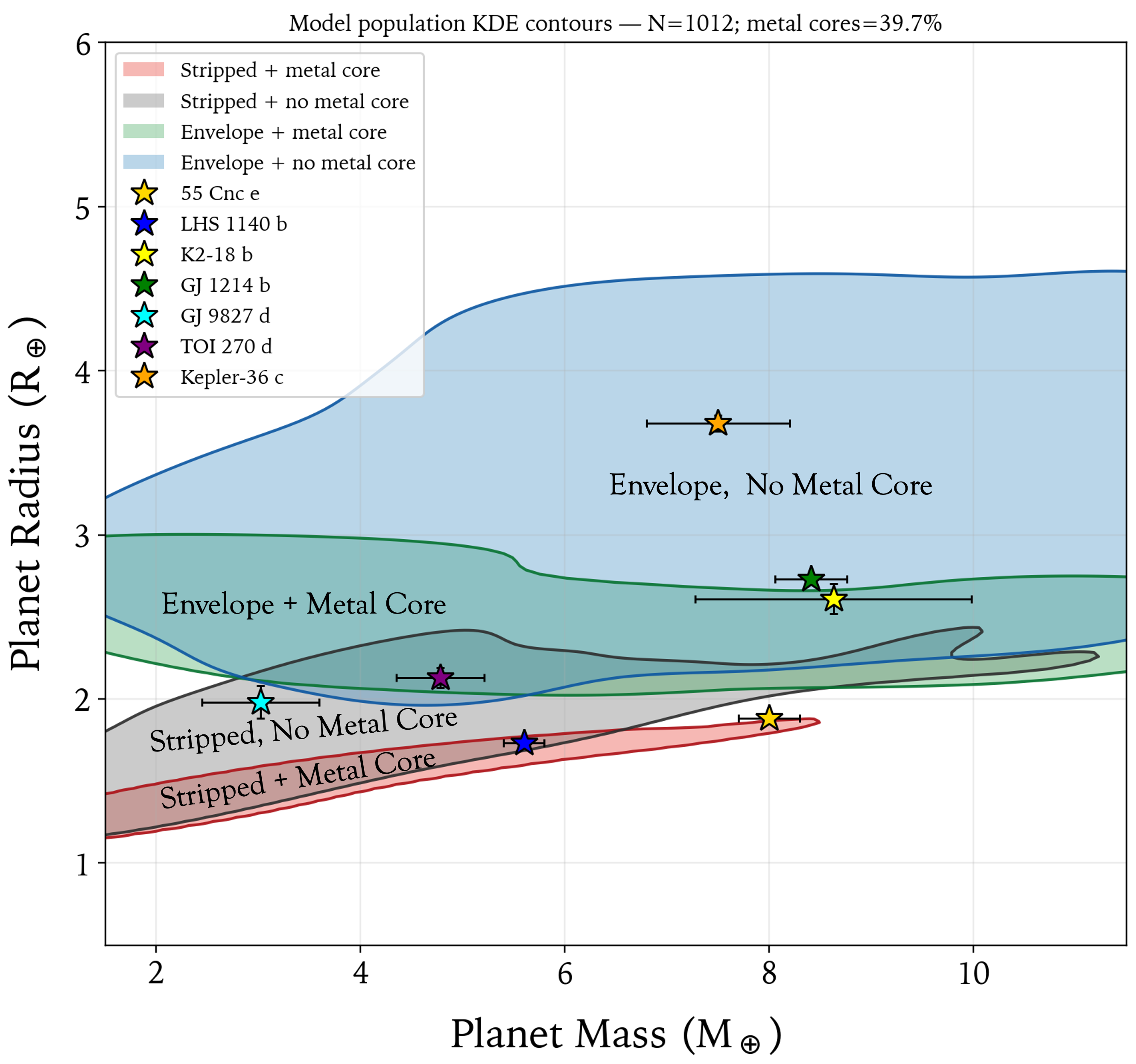}
    \caption{Comparison of fields for planet types defined by model planets with observed planets in mass vs.\ radius space. The regions associated with the four model groups are defined by KDE contours enclosing locations where the density exceeds 5\% of the peak value for that group. The four groups are: 1) differentiated planets with metal cores stripped of their H$_2$-rich atmospheres; 2) planets with no discrete metal cores and stripped of their H$_2$-rich atmospheres; 3) differentiated planets with metal cores that retain their H$_2$-rich envelopes, and 4) planets with no metal cores that retain their H$_2$-rich envelopes. }
\label{fig:Mass_vs_Radius_fields}
\end{figure}

Despite these shortcomings, the model planet types appear to cluster in ways reminiscent of the different observed sub-Neptunes and super-Earths. It is therefore tempting to try to relate the variety of planets exemplified by the examples described in \S \ref{section:reference_planets} and shown in Figure \ref{fig:Mass_vs_Radius_REFERENCE} to the models shown in Figure \ref{fig:Mass_vs_Radius}. To facilitate this comparison, the mass-radius fields defined by the four types of model planets together with the example observed planets are shown in Figure \ref{fig:Mass_vs_Radius_fields}. The fields are defined by KDE contours for each group.  Although these fields exhibit overlaps, they are nonetheless useful for guiding inferences about the nature of planets based on their bulk densities in the context of hydrogen-silicate-iron miscibility.

Planets 55 Cancri e and LHS-1140 b plot in the ``Stripped + Metal Core" field in Figure \ref{fig:Mass_vs_Radius_fields}. This fields corresponds to the white diamonds with red dots in Figure \ref{fig:Mass_vs_Radius}, representing differentiated bodies stripped of their primary atmospheres.  These planets are basically pure rock and metal but with density deficits. The models here would suggest that the low densities of these planets compared with pure rock and metal, molten or solid, could be explained by significant mass fractions of hydrogen in their interiors (in whatever form).  This is not unlike Earth, which is underdense relative to pure rock and iron metal due to light elements, including H, in the metal core \citep{Schlichting_Young_2022, Young_Nature_2023}.

Planets GJ 9827 d and TOI 270 d plot in the ``Stripped, No Metal Core" and ``Envelope + Metal Core" fields in Figure \ref{fig:Mass_vs_Radius_fields}, representing the black diamonds and circles with red dots in Figure \ref{fig:Mass_vs_Radius}, respectively. Recall that GJ 9827 exhibits evidence for a steam-rich atmosphere, suggesting it may have lost its H$_2$-rich envelope.  This is consistent with its plot location in mass-radius space nearer to the "Stripped, No Metal Core" field.  TOI 270 d, on the other hand, is thought to have retained an H$_2$-rich envelope modified by interaction with a magma ocean.  This is consistent with its plotting within the ``Envelope + Metal Core" field in Figure \ref{fig:Mass_vs_Radius_fields}. 
The models provide complementary, and not necessarily orthogonal, explanations for planets with broadly similar bulk densities: their cores may be undifferentiated and enriched in hydrogen, reducing the mean interior density without requiring massive hydrogen-rich envelopes. The model planets here do not preclude secondary, steam atmospheres in these cases provided that such atmospheres do not significantly inflate the planetary radii.

Planets GJ 1214 b and K2-18 b fall squarely in the peak of the sub-Neptune probability density in mass-radius space, straddling the boundary between the fields that both correspond to extant H$_2$-rich envelopes but with or without discrete metal cores in Figure \ref{fig:Mass_vs_Radius_fields}. They therefore most resemble model planets with substantial H$_2$-rich envelopes. However, recall that while K2-18 b is thought to have a hydrogen-rich envelope, GJ 1214 b may not. We note that the ``Stripped, No Metal Core" field in Figure \ref{fig:Mass_vs_Radius_fields} extends towards these two planets, perhaps providing an explanation for GJ 1214 b not having a hydrogen-rich envelope. 

Planet Kepler-36 c resembles the model planets with substantial H$_2$-rich envelopes with mass fractions up to $\sim 4\%$ by mass and no metal cores (Figure \ref{fig:Mass_vs_Radius_fields}). The models produce a significant range in H$_2$ atmosphere mass fractions at similar positions in mass-radius space (Figure \ref{fig:Mass_vs_Radius}). 

\begin{figure}[t!]
\centering
   \includegraphics[width=0.46\textwidth]{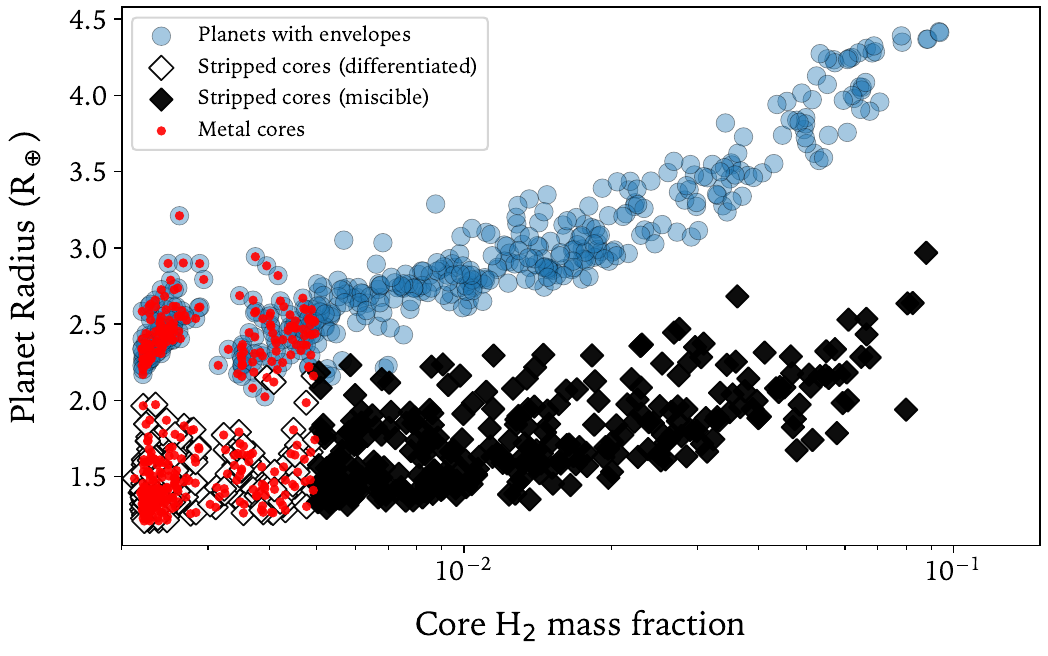}
    \caption{Plot of model planet radii versus the  mass fraction of H2 in their interiors.  Symbols are the same as those used in Figure \ref{fig:Mass_vs_Radius}. The sharp edge to the distribution at low mass fractions of H$_2 \sim 0.002$ is a reflection of the lower limit for numerically stable H$_2$ atmospheres in the models.}
\label{fig:Radius_vs_coreH2}
\end{figure}

The intrinsic degeneracies in mass-radius space that remain reflect the significant effects of large fractions of hydrogen in interiors can have on stripped cores, as shown in Figure \ref{fig:Radius_vs_coreH2}.  Our models underscore this effect in a systematic way.

\subsection{Assumptions and Caveats}
In our inverse epistemological approach, we have gone to some effort to reduce assumptions about the formation of the model planets by, for example, mitigating the need to enforce initial intrinsic luminosities and justifying the prior distributions for input parameters with evolution models. However, a parameter not yet discussed in detail so far is the median bulk H$_2$ mass fraction prior to atmospheric escape used for our random draws. This parameter is, in effect, tuned to yield radii similar to the observed planets. 

We show in Appendix B a simple calculation suggesting that accretion accompanied by simultaneous fractional escape of hydrogen should lead to a log-normal distribution like that used in our models for reasonable estimates of parameters.  In this simple calculation, the median value for planet mass fractions of H$_2$ of $1.8\%$ corresponds to accretion of protoplanetary bodies ranging in mass from $0.1$ to $2.0 M_\oplus$, with each accretion event resulting in retention of between 18\% and 100\% of primary atmospheres. The retention factor for each event is drawn at random from a parent population (Appendix B). The fact that the median mass fraction of H$_2$ for our model planets fits both the radius valley and the radius ``cliff" (the drop off in radii at about $4 R_\oplus$) provides some {\it a posteriori} confidence in its significance. A more thorough modeling of the accretion process is warranted. 

The relevant question raised by the initial
condition implicit in this study is whether accretion, including giant impacts in particular, results in intermingling of hydrogen and molten rock.  Several lines of evidence indicate that mechanical and chemical co-mingling of hydrogen and silicate melt is likely.
Three-dimensional smoothed-particle hydrodynamics (SPH) simulations show that
individual giant impacts involving Earth-mass bodies rarely remove more than
${\sim}10\%$ of H$_2$--He primary envelopes comprising $5\%$ of the targets, with cumulative losses over many impacts reaching $40$ to $70\%$ \citep{Roche2025}; large fractions of primary hydrogen accreted by protoplanets survive the violent accretion process. SPH typically suppresses material mixing across steep density contrasts and so cannot resolve fine-scale entrainment of gas into melt. Nonetheless, \citet{Roche2025} infer that surviving volatiles are mostly dissolved in the silicate.
\emph{Ab initio} molecular dynamics simulations show that giant impacts drive
portions of the colliding bodies to supercritical states, resulting in a continuous liquid--vapor structure
\citep{CaracasStewart2023}. In the resulting impact-generated ``synestia", volatiles are well mixed with the more refractory material \citep{Lock2018}. High-velocity impact simulations further show vigorous mechanical mixing of  primary envelope gas and vaporized rock followed by hydrogen--rock melt mixing \citep{KurosakiInutsuka2023,
Kurosaki2023}. More work is required on the fate of hydrogen during accretion of protoplanetary materials, but thus far,  substantial hydrogen in planet interiors seems to be a viable initial condition.

The connection to observations in this study is limited to comparisons in mass–radius and radius–period space and is therefore based on bulk density constraints alone. The physical framework is restricted to hydrogen–silicate–iron phase equilibria and does not explicitly include additional volatile elements. Consequently, the models do not provide quantitative predictions for atmospheric compositions or spectral signatures. A direct comparison with transmission or emission spectra would require coupling the interior structure calculations to a more comprehensive chemical equilibrium framework such as \citet{Schlichting_Young_2022} linked self-consistently to an atmospheric structure model \citep[e.g.,][]{Nixon2025_TOI270d}. Establishing this connection between interior phase equilibria and observable atmospheric chemistry remains an important next step.

\section{Conclusions}
\label{section:conclusions}
Model planets based on the concept of variable miscibility between hydrogen, molten silicate, and molten iron metal, together with atmosheric escape, can reproduce roughly 80\% of the occurrence density structure of sub-Neptunes and super-Earths in mass vs.\ radius space. The model planet realizations also mimic the radius valley and radius vs.\ period distributions of the actual planets. This overlap in predicted occurrences with observations suggests that hydrogen-silicate-iron miscibility 
may serve as a unifying concept for planet formation for sub-Neptunes and super-Earths.  In these models, the boundary between supercritical magma oceans and overlying hydrogen-rich envelopes is the binodal that specifies the equilibrium condition between the condensed phase and the atmosphere/envelope.  

Planets modeled this way may explain the full range of planet types exhibited by sub-Neptunes and their potential super-Earth descendents.  A key outcome is the distribution of hydrogen between planet interiors and their outer envelopes.  Planets with hydrogen $< \sim 1\%$ by mass when they formed produce discrete, terrestrial-like metal cores, while planets born with more hydrogen are predicted to have interiors composed of a single phase due to the effects of hydrogen. While necessarily incomplete, hydrogen-silicate-iron miscibility offers an overarching explanation for the wide variety of planet types among sub-Neptunes and super-Earths based solely on the accreted mass fractions of hydrogen-rich primary atmospheres and, importantly, the phase equilibria involving silicate, Fe metal, and H$_2$.

\section*{Appendix A}
\label{section:Appendix_A}

\begin{center}
\textit{Data Extracted from the NASA Archive}
\end{center}
\smallskip

The exoplanet sample used in this work was obtained from the NASA Exoplanet Archive with the Table Access Protocol (TAP) synchronous interface, querying the \texttt{pscomppars} table. Our query retrieved planet mass, radius, orbital period, mass provenance flags, and host-star properties including stellar mass and stellar age with associated uncertainties. We selected planets with reported masses within a specified interval from $1.5$ to $11.5$ $M_\oplus$, in keeping with our model mass range. We set a maximum orbital period of 100 days. Planets were further required to be associated with usable uncertainty estimates, meaning both $M_p$ and $R_p$ must have reported one-sided uncertainties. A quality cut was applied based on fractional mass precision,
\begin{equation}
\frac{\sigma_M}{M_p} \le \epsilon_M = 0.3.
\end{equation}
In order to account for asymmetric errors, the adopted fractional uncertainty is taken to be 
\begin{equation}
\sigma_M = \max\!\left(|\sigma_{M,+}|, |\sigma_{M,-}|\right),
\end{equation}
so that the most conservative reported uncertainty was used.

In the present work we include all
observationally determined masses reported in the archive, including
measurements derived from radial-velocity (RV) analyses and
transit-timing variations (TTVs), while excluding masses inferred
from empirical mass--radius relations; only dynamically or observationally measured planet masses are
retained. 

\begin{center}
\textit{Pseudo-completeness Correction for the Radius Distribution}
\label{app:completeness}
\end{center}
\smallskip

A completeness-corrected planet-radius distribution
was created to enable comparison between observed exoplanet samples and model
populations. The method provides an approximate occurrence-like
quantity (planets per star per radius bin) without performing a full
injection--recovery analysis.

Planet properties $(R_p, P)$ and host identifiers are obtained from the
NASA Exoplanet Archive along with their corresponding host star properties, including 
 radius $R_\star$,  mass $M_\star$, observing baseline
$T_{\rm obs}$, duty cycle $d$ (the fraction of the observing baseline
during which photometric data were collected), and the Combined
Differential Photometric Precision \citep{Christiansen2012},
a per-star photometric noise metric evaluated at a reference transit
duration.  These quantities define detectability for hypothetical
planets around each target star.

Planets are grouped into radius bins $\mathcal{B}_j$.
Each bin is assigned a representative radius $R_j$ and a
representative orbital period $P_j$, taken as the median period of
planets within the bin.  This preserves the dependence of detection
efficiency on orbital period while maintaining a one-dimensional
radius distribution.

For a planet of radius $R$ and period $P$ orbiting a star,
the transit depth is

\begin{equation}
\delta = \left(\frac{R}{R_\star}\right)^2 .
\end{equation}
Assuming circular central transits, the duration is approximated by
\begin{equation}
T_{14} \approx \frac{P}{\pi}\frac{R_\star}{a}, \qquad
a = \left(\frac{G M_\star P^2}{4\pi^2}\right)^{1/3},
\end{equation}
where $T_{14}$ denotes the total transit duration, defined as the
interval between first and fourth contact, i.e., when the
planetary and stellar limbs first and last become tangent.
The expected number of observed transits is
\begin{equation}
N_{\rm tr} = d\,\frac{T_{\rm obs}}{P}.
\end{equation}

From these values we obtain the photometric noise scaled with transit duration as
$\sigma(T_{14}) \propto T_{14}^{-1/2}$, yielding an approximate
signal-to-noise ratio accumulated over multiple transit events,
i.e., the multiple-event statistic \citep{Jenkins2002},
\begin{equation}
\mathrm{MES}(R,P|s)
\simeq
\frac{\delta \sqrt{N_{\rm tr}}}{\sigma(T_{14})}.
\end{equation}

Detection efficiency can be modeled using a sigmoidal threshold function based on the MES
\citep{Burke2015}:
\begin{equation}
\epsilon(\mathrm{MES})
=
\frac12
\left[
1+\mathrm{erf}
\left(
\frac{\mathrm{MES}-\mu}{\sqrt{2}\sigma}
\right)
\right],
\end{equation}
with $(\mu,\sigma)=(8,1.5)$, values calibrated to the
\textit{Kepler} pipeline detection threshold \citep{Jenkins2010,Burke2015}.

For each radius bin $\mathcal{B}_j$, the effective number of searchable
stars is
\begin{equation}
D_j =
\sum_s
\mathbf{1}\!\left[N_{\rm tr}(P_j|s)\ge N_{\rm tr,min}\right]
\,
\epsilon\!\left(\mathrm{MES}(R_j,P_j|s)\right),
\end{equation}
where $N_{\rm tr,min}=3$ is the minimum number of observed transits
required for a detection threshold consistent with the \textit{Kepler}
pipeline \citep{Jenkins2010}.

If $\mathcal{P}_j$ denotes planets with radii within bin
$\mathcal{B}_j$, the occurrence per star per bin is
\begin{equation}
f_j = \frac{|\mathcal{P}_j|}{D_j}.
\end{equation}
We note that this expression does not include a correction for the
geometric transit probability $p_{\rm tr} = R_\star/a$; the quantity
$f_j$ should therefore be interpreted as a detection-efficiency-corrected
count rather than a true occurrence rate \citep[cf.][]{Howard2012,Fressin2013}.
In practice, each planet is assigned a weight
\begin{equation}
w_{\rm occ} = \frac{1}{D_j},
\end{equation}
and the radius distribution is obtained by summing these weights within
bins. Summing the weights of planets within a bin reproduces the
occurrence estimate,
\begin{equation}
\sum_{i \in \mathcal{P}_j} w_{\rm occ} = f_j .
\end{equation} 

\vspace{\baselineskip}
\section*{Appendix B}
\label{section:Appendix_B}

\begin{center}
\textit{Hydrogen Accretion }
\end{center}
\smallskip

The observed spread in inferred H$_2$ mass fractions among sub-Neptunes is well described by the log-normal distribution used in our models with a median value of $1.8\%$ by mass. In lieu of a detailed and highly uncertain accretion model, here we describe a simple stochastic numerical accretion experiment that provides a physical motivation for such a distribution. The goal is not to construct a full accretion model, nor to fit the adopted distribution perfectly, but rather it is to demonstrate that log-normal distributions of H$_2$ fractions arise naturally from the combination of discrete accretion events and partial losses of envelopes surrounding the accreting bodies in general. 

In the experiment, planets are assembled by sequential addition of smaller solid bodies drawn at random from a prescribed mass distribution until a target core mass is reached. Each accreted body has mass $m_i$, and the total solid mass is the sum of these contributions. During a sequence of accretion events, two processes occur recursively. First, the impact associated with each accretion event partially strips the pre-existing H$_2$ envelope; after event $i$, only a fraction $x_{r_i}$ of the envelope mass present prior to the next collision survives. Second, the newly accreted body contributes additional H$_2$ according to the cooling-limited scaling of \cite{ginzburg2016a}, in which the envelope mass fraction associated with a body of mass $m$ scales approximately as
\begin{equation}
f=
0.02
\left(\frac{m}{M_\oplus}\right)^{0.8}
\left(\frac{T}{1000\,{\rm K}}\right)^{-1/4}
\left(\frac{t_{\rm disk}}{1\,{\rm Myr}}\right)^{1/2},
\end{equation}
where $M_\oplus$ is one Earth mass, $T$ is the characteristic disk temperature at the planet's location, and $t_{\rm disk}$ is the gas disk lifetime.  The increment in the growing H$_2$-rich envelope mass during each accretion event is then $\Delta M = f(m)\, m$. The envelope mass therefore evolves according to
\begin{equation}
M_{\rm env}^{(k+1)} = x_{r_k} M_{\rm env}^{(k)} + \Delta M_k,
\label{eqn:Mk}
\end{equation}

\noindent where $\Delta M_k$ is the increment of H$_2$ accreted during event $k$. The final H$_2$ mass fraction is then

\begin{equation}
f_{\rm H_2} = \frac{M_{\rm env}}{M_{\rm core} + M_{\rm env}}.
\end{equation}

The essential statistical feature of the model is the multiplicative survival factor $x_{r_k}$. Physically, $x_{r_k}$ represents the combined effects of impact-driven stripping, shock heating, transient hydrodynamic escape, or variations in cooling efficiency. We envision the total retention factor during a given accretion event schematically as a product of independent contributions. Provided that these individual contributions have finite variance and are not dominated by a single catastrophic process, the central limit theorem implies that the logarithm of the net retention factor is approximately Gaussian in its distribution.

Iterating Equation \ref{eqn:Mk} through N accretion events gives
\begin{equation}
M_{\rm env}^{(N)} =
\left(\prod_{j=0}^{N-1} x_{r_j}\right) M_{\rm env}^{(0)}
+
\sum_{k=0}^{N-1}
\left[
\Delta M_k
\prod_{j=k+1}^{N-1} x_{r_j}
\right],
\label{eqn:MN_iterated}
\end{equation}
so that the final envelope mass is the sum of the initial envelope mass and all subsequent accretion increments, each weighted by the product of retention fractions associated with later impacts.

While $M_{\rm env}$ is not a pure product, its value is strongly influenced by accumulated multiplicative retention factors. When many stochastic accretion events contribute comparably and no single increment dominates the total budget, the products $\prod x_{r_j}$ introduce effectively random  variability to the hydrogen accretion process. Under these conditions, $\ln M_{\rm env}$ behaves approximately as a sum of random contributions associated with $\ln x_{r_j}$ and modestly varying accretion terms. The distribution of $M_{\rm env}$ consequently approaches a log-normal distribution. Since the final core mass is fixed by construction in this experiment, the H$_2$ mass fraction inherits this approximately log-normal distribution.
The log-normality is not imposed \emph{a priori}. Rather, it emerges from the combined effects of stochastic growth and partial loss of hydrogen during accretion events. 
\begin{figure}
\centering
   \includegraphics[width=0.43\textwidth]{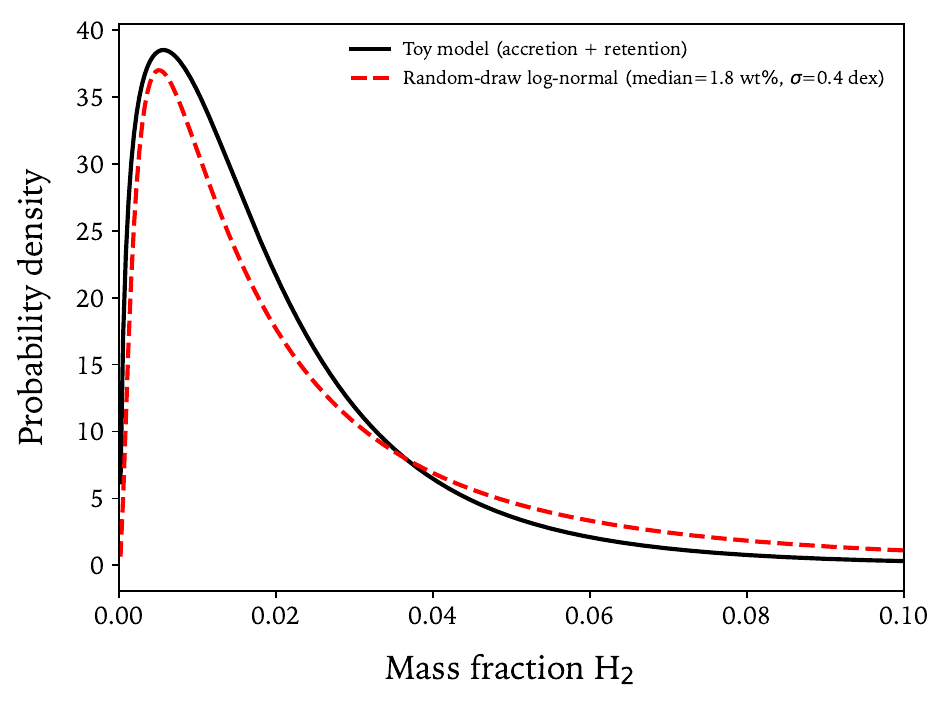}
    \caption{Distribution of hydrogen mass fractions produced by the simple accretion calculation for 30,000 planets each with a total mass of $5 M_\oplus$. The distribution used to generate the model planets in this study is shown for comparison (dashed curve).  }
\label{fig:simple_H2_distribution}
\end{figure}

Figure \ref{fig:simple_H2_distribution} shows results from this calculation for a range of accreting masses from $0.1$ to $2.0 M_\oplus$ and an envelope retention fraction $x_{r_i}$ drawn from a log-normal distribution defined by $\ln r \sim \mathcal{N}(\mu_{\log}=0.8, \sigma_{\log}=1.5)$, corresponding to a range of retention fractions from $0.18$ to the maximum value of unity. The curve resembles the distribution used to generate the model planets (Figure \ref{fig:simple_H2_distribution}.

This simple calculation illustrates that a log-normal distribution of H$_2$ mass fractions can arise naturally as a statistical consequence of hierarchical accretion combined with incremental losses of hydrogen for a reasonable set of parameters. In this sense, the adopted log-normal description of sub-Neptune H$_2$ mass fractions in our models can be viewed as an emergent property of stochastic assembly.

\section*{Acknowledgments} 
We acknowledge financial support from NASA grant 80NSSC24K0544 (Emerging Worlds program).  We are indebted to James G. Rogers (Institute of Astronomy, University of Cambridge) who contributed to the  interior model Python code used in this study.

\bibliographystyle{aasjournal}
\bibliography{edy_references}

\end{document}